\documentclass[oldversion]{aa}
\usepackage{graphicx}
\usepackage{aalongtable}
\usepackage{txfonts}
\usepackage{rotating}
\usepackage{lscape}
\newcommand{\gsim}{\;\lower.6ex\hbox{$\sim$}\kern-7.75pt\raise.65ex\hbox{$>$}\;}
\newcommand{\lsim}{\;\lower.6ex\hbox{$\sim$}\kern-7.75pt\raise.65ex\hbox{$<$}\;}

\begin{document}
\title{Na-O Anticorrelation and HB. VIII. Proton-capture elements and
metallicities in 17 globular clusters from UVES 
spectra\thanks{Based on observations  collected at ESO telescopes under
programmes 072.D-507 and 073.D-0211}\fnmsep\thanks{
   Tables 1, 3, 4, 6, 8, and 9 are only available in electronic form at the CDS via anonymous
   ftp to {\tt cdsarc.u-strasbg.fr} (130.79.128.5) or via
   {\tt http://cdsweb.u-strasbg.fr/cgi-bin/qcat?J/A+A/???/???}}
 }

\author{
Eugenio Carretta\inst{1},
Angela Bragaglia\inst{1},
Raffaele Gratton\inst{2},
\and
Sara Lucatello\inst{2,3}
}

\authorrunning{E. Carretta et al.}
\titlerunning{Proton-capture elements in 17 globular clusters}

\offprints{E. Carretta, eugenio.carretta@oabo.inaf.it}

\institute{
INAF-Osservatorio Astronomico di Bologna, Via Ranzani 1, I-40127
 Bologna, Italy
\and
INAF-Osservatorio Astronomico di Padova, Vicolo dell'Osservatorio 5, I-35122
 Padova, Italy
\and
Excellence Cluster Universe, Technische Universit\"at M\"unchen, 
 Boltzmannstr. 2, D-85748, Garching, Germany 
  }

\date{Received 18 March 2009 / Accepted 31 May 2009}

  \abstract{We present homogeneous abundance determinations for iron and some
  of the elements involved in the proton-capture reactions (O, Na, Mg, Al, and Si)
  for 202 red giants in 17 Galactic globular clusters (GCs) from the analysis of 
  high-resolution UVES spectra obtained with the FLAMES facility at the ESO VLT2 
  telescope. Our programme clusters span almost the whole range of the metallicity
  distribution of GCs and were selected to sample the widest range
  of global parameters (horizontal-branch morphology, masses, concentration,
  etc). In this paper we focus on the discussion of the Na-O and Mg-Al 
  anticorrelations and related issues. Our study finds clear Na and O 
  star-to-star abundance variations, exceeding
  those expected from the error in the analysis, in all clusters. Variations in
  Al are present in all but a few GCs. Finally, a spread in abundances of
  Mg and Si are also present in a few clusters. Mg is slightly less overabundant
  and Si slightly more overabundant in the most Al-rich stars. 
  The
  correlation between Si and Al abundances is a signature of production of
  $^{28}$Si leaking from the Mg-Al cycle in a few clusters. The 
  cross sections required for the proper reactions to take over in the cycle 
  point to temperatures in excess of about 65 million K for the favoured site of
  production. We used a dilution model to infer the total range of Al abundances 
  starting from the Na and Al abundances in the FLAMES-UVES spectra, and the Na 
  abundance distributions found from analysis of the much larger set of stars for 
  which FLAMES-GIRAFFE spectra were available. We found that the maximum amount 
  of additional Al produced by first-generation polluters contributing to the
  composition of the second-generation stars in each cluster is closely correlated 
  with the same combination of metallicity and cluster luminosity that reproduced 
  the minimum O-abundances found from GIRAFFE spectra. We then suggest that the high 
  temperatures required for the Mg-Al cycle are only reached in the most massive 
  and most metal-poor polluters.
}

\maketitle

\section{Introduction}

A growing body of evidence indicates that globular clusters (GCs) are not actual
examples of simple stellar populations: when examined in detail,  their stars
are not coeval, nor do they share the same initial  chemistry (see the
exhaustive reviews by Kraft 1994 and, more recently, by  Gratton, Sneden,
Carretta 2004 on this subject). The star formation in  GCs included a second
generation of stars born from the ejecta of the most  massive among the stars of
the primeval generation. The first phases of  evolution of GCs were likely
complex, including a variety of dramatic and  energetic phenomena like supernova
explosions and high and low velocity winds  from both blue and red massive stars,
possibly combining in giant expanding bubbles  of gas and shock fronts, which
may have triggered further star formation.  Even more complex might have been
the environment for these events, possibly  in the cores of giant clouds or
even, in the case of the most massive clusters,  of dwarf galaxies (Bekki 2006).

Most of these events left behind a trace, represented by the chemical
composition of the stars, that can be followed through quite subtle spectral
features, like the lines of O, Na, and another few key elements such as Al, Mg,
and possibly Si. This allows us to quantitatively probe the existence 
of a  second generation of stars in GCs.

The best known among these tracers is the star-to-star anticorrelation between 
the O and Na abundances, which has been found in all GCs surveyed so far, at 
variance with field stars (see e.g. Gratton et al. 2000). This is the sign of 
(originally unexpected) material processed through the complete 
CNO and the Ne-Na cycles of proton capture reactions in -possibly-
the majority of GC stars.  The key study of non-evolved stars in GCs by Gratton
et al. (2001) unambiguously  showed that the  processed material must
originate in an earlier generation  of stars (see also Cohen et al. 2002).
Self-pollution models are not yet able to  convincingly explain all observed
features. Candidate first-generation polluters are thermally pulsing,
intermediate-mass AGB stars undergoing hot bottom burning (see Ventura et al.
2001), or massive, rotating stars (see Decressin et al. 2007). Addition of new
observational data is crucial to better constraining these models. 

In addition to the CNO and Ne-Na cycle, there is evidence that the Mg-Al
cycle is also active in GC polluters.
The Mg-Al cycle provides information that is complementary to those provided by
the Ne-Na cycle,  mainly for two reasons: $(i)$ it requires much higher
temperatures (some 70 MK to be compared with 25 MK, see Charbonnel \& Prantzos
2006), meaning that different polluters are possibly involved; $(ii)$ while the
Ne-Na cycle essentially saturates in most polluters, thus providing similar
maximum Na  abundances in most GCs, with only small (though detectable, see
below) variations, no case of complete transformation of Mg into Al has been
observed. The case of  the Mg-Al is clearly more complex, because many nuclei
are involved (the various  Mg isotopes; see e.g. Yong et al. 2003), with
possibly  even some production of Si  (see e.g. Yong et al. 2005). Hence, while an
overall correlation between Na and Al  might be expected (because they should
share dilution effects like those considered by Prantzos \& Charbonnel 2006),
details of this correlation are not obvious and can provide crucial information
on the nature of the polluters.

In addition to modifications in the distribution of products from H-burning at 
high temperature, GCs could significantly differ from the field population if 
the proto-GCs were able to have some independent chemical evolution, retaining 
at least part of the metal-enriched ejecta of core-collapse SNe (e.g, Cayrel
1986;  Parmentier \& Gilmore 2001). Observationally, we might expect that GCs
have a higher  [$\alpha$/Fe]\footnote{We adopt the usual spectroscopic notation,
$i.e.$ [X]=  log(X)$_{\rm star} -$ log(X)$_\odot$ for any abundance quantity X,
and log  $\epsilon$(X) = log (N$_{\rm X}$/N$_{\rm H}$) + 12.0 for absolute
number density  abundances.} ratio than field stars of similar metal abundance,
as well as possibly  other peculiar abundance patterns (concerning e.g.
primordial Al and Na abundances).  These signatures are likely subtle and might
have escaped detection in literature data, lacking in homogeneity.

In this series of paper we are presenting a very extensive new set of data on
abundances in several Galactic GCs, selected to span all the  ranges
of different physical parameters (metallicity, concentration, density,  HB
morphology, mass, etc.), obtained exploiting the ESO multi-fibre FLAMES 
facility. This survey (see Carretta et al. 2006a, hereafter Paper I) aims to
provide significant constraints and answers to fundamental questions  such as:
{\it (i)} Were GC stars really born in a single ''instantaneous''  burst? {\it
(ii)} Did GCs self-enrich? {\it (iii)} How do abundance anomalies  within each
individual GC relate to the formation and early evolution of the  GC itself and
of each individual member? 

Using FLAMES/GIRAFFE spectra of more than 2000 stars in 19 GCs we have singled
out in  a companion paper (Carretta et al. 2009, hereinafter Paper VII)
second-generation stellar populations in many GCs. Using the Na-O
anticorrelation as a diagnostic,  we studied the enrichment history in GCs. The
main results of that paper were that 
\begin{itemize}
\item [$i)$]  the Na-O anticorrelation is present
in all GCs and the second generation stars (that we separate into an intermediate
and extreme I and E populations) dominate the first generation stars (that
we called primordial P population); 
\item [$ii)$] the E and P populations are more
numerous among the most massive clusters; 
\item [$iii)$] the shape of the Na-O
anticorrelation changes regularly (from more extended and shallow to less
extended and steep) as a  function of cluster metallicity and luminosity (this
last a likely proxy  for the average polluter mass); 
\item [$iv)$] the amount
of Na produced in more  luminous and metal-poor clusters is close to what is
expected from simple  transformation of all $^{22}$Ne originally available in
the polluters into  $^{23}$Na, while an
extra-amount of  $^{22}$Ne (possibly due to primary production from
triple-$\alpha$ burning) is  required in less luminous and metal-rich clusters.
\end{itemize}

In this paper we present the analysis of the high-resolution UVES spectra  that
we collected simultaneously with the moderate resolution GIRAFFE  spectra, a 
unique and precious opportunity provided by the FLAMES multiobject facility.
While only seven fibres can be devoted to UVES in each pointing (more than  an
order of magnitude less than for GIRAFFE), UVES spectra have a higher resolution
and  ten times wider spectral coverage. For this reason they were crucial in
several respects:
\begin{itemize}
\item [1)] to measure accurate O and Na abundances for additional stars beside
those with GIRAFFE spectra; this point turned out to be important in some cases
(e.g. NGC~6397) where O abundances  could be derived only for a handful of
stars  from lower resolution GIRAFFE observations (see Paper VII);
\item [2)] to provide accurate measurements of equivalent widths ($EW$s) from
high-resolution spectra to build a reference system on which to check the
$EW$s measured on lower resolution GIRAFFE spectra for each cluster;
\item [3)] to obtain abundances for the proton-capture element Al with no
transition falling in the spectral range covered by the two GIRAFFE gratings
used (HR11 and HR13);
\item [4)] to define a new metallicity scale based on accurate measurements in a
large sample of clusters, homogeneously derived with modern sets of near
infrared magnitudes (from 2MASS, Skrutskie et al. 2006), atmospheric parameters
(from the calibration by Alonso et al. 1999), and atomic parameters (from the
compilation by Gratton et al. 2003);
\item [5)] to test the idea by Cayrel (1986) that proto-GCs might have been able
to sustain some independent chemical evolution, aside from modification in CNO
elements, retaining at least part of the metal-enriched ejecta of core-collapse
supernovae. 
\end{itemize}

Previous papers of this project have focused on a few particular clusters, tuning our
procedures for the analysis of huge samples of stars and for estimates of error
sources, especially from GIRAFFE spectra. In the companion Paper VII we
presented the Na-O anticorrelation derived for almost 2000 red giant branch
(RGB) stars and provided a synoptic table with references to all the analyses.
 UVES spectra were only exploited up to now to measure the detailed
chemical composition  of two poorly studied, yet interesting bulge clusters
(NGC~6441, Gratton et al. 2006,  Paper III; and NGC~6388, Carretta et al. 2007b,
Paper VI), for a total of 12 stars. In the present paper we  present the
analysis of the other 202 RGB star with  UVES spectra that are {\it bona fide}
members in the remaining 17 clusters of our  sample. We  focus our
discussion on the additional information they can provide  about the proton
capture process (the first three items listed above), while an in depth 
analysis of the information concerning primordial abundances as well as
star-to-star  scatter in Fe abundances will be presented in future works.   The 
most crucial additional data concern the Mg-Al cycle, which could not be observed
on the GIRAFFE spectra for lack of Al lines. (We selected the two gratings
including the O and Na spectral features most suited for abundance 
determination).

The outline of the paper is as follows. Criteria of target selection and
observations are given in the next section; our procedure deriving the
atmospheric  parameters and the analysis are described in Sect. 3, whereas error
estimates are briefly discussed  in Sect. 4.  In Sect. 5 we show the pattern of
Na-O and Mg-Al anticorrelations in all  clusters, while Sect. 6 is devoted to
extending and tuning the simple dilution model set up in the companion Paper VII,
with a comparison with predictions from models of massive stars (both in main
sequence and in the AGB phase). Finally, results are summarised in  Sect. 7.

\section{Target selection and observations}

Our observations were collected in Service Mode at the ESO VLT-UT2 telescope 
under programmes 072.D-0507 and 073.D-0211 using the high-resolution multifibre
facility FLAMES (Pasquini et al. 2002). In Table~\ref{t:log} we list the log of
observations for each cluster; date and time of beginning of the observations
are UT, exposure times are in seconds. The reported airmass is the one at the
start of each exposure.
The UVES Red Arm fed by FLAMES fibres provides a wavelength coverage from 4800
to 6800~\AA\ with a resolution of $R\simeq 40,000$.

\begin{table}
\centering
\scriptsize
\caption{Log of the UVES observations (complete table available only in
electronic form}
\begin{tabular}{lccrl}
\hline
 GC      &    date   & UT          & Texp&airmass \\
\hline
NGC0104  &2004-06-26 &09:47:32.114 &1600 &1.497 \\
         &2004-07-07 &09:09:12.509 &1600 &1.494 \\
         &2004-07-07 &09:47:22.887 &1600 &1.478 \\
\hline
\end{tabular}
\label{t:log}
\end{table}

\paragraph{Cluster selection-} The first criterion
adopted was to select clusters with a wide range of distribution of stars along
the HB, since this project mainly focused on 
studying  possible relations between the morphology of the HB and the chemical
composition of different stellar generations in clusters. 
In our  sample all HB morphologies are represented, from the stubby red type to
clusters with only blue  HBs and with very extended blue tails, passing through
bimodal or red and blue HB  morphologies. Second, since the
metallicity is the well-known  {\it first} parameter in shaping the HB
morphology, we inserted GCs  with metal abundances from
[Fe/H]$=-2.4$\ to about [Fe/H]$=-0.4$ in our sample, spanning almost  the whole metallicity
range of the Galactic GCs. Both small and  massive clusters are
included in our sample, and finally all the age classes defined  by Zinn, Lee,
and
collaborators (see e.g. Mackey and van den Bergh 2005) are  represented well in
the sample (see also Carretta et al. 2009, Paper VII). In  Table\ref{t:snrv} we
provide some information for the sample analysed here. After the  cluster
alternate name we list the number of member stars observed, together with  their
average effective temperature and the mean $S/N$  per pixel, estimated 
around 6200~\AA. 

\begin{table*}
\centering
\caption{Information on the observed GCs}
\begin{tabular}{rlrccccc}
\hline
 GC      & other & Nr$_{\rm stars}$ & $<T_{eff}>$ & $<S/N>$ & $RV$ & $\sigma$ & $RV_{Harris}$ \\   
\hline
NGC ~104  & 47 Tuc  & 11 &4125 & 95 & -19.86 & 7.26 & -18.7\\
NGC ~288  &         & 10 &4465 &101 & -46.15 & 2.55 & -46.6\\
NGC 1904  & M 79    & 10 &4520 & 69 & 204.82 & 2.02 & 206.0\\
NGC 2808  &         & 12 &4423 & 97 & 104.09 & 8.01 &  93.6\\
NGC 3201  &         & 13 &4441 & 83 & 494.57 & 3.85 & 494.0\\  
NGC 4590  & M 68    & 13 &4668 & 86 & -94.35 & 2.24 & -94.3\\ 
NGC 5904  & M 5     & 14 &4438 & 70 &  53.78 & 5.91 &  52.6\\ 
NGC 6121  & M 4     & 14 &4329 & 91 &  71.95 & 4.07 &  70.4\\ 
NGC 6171  & M 107   &  5 &4427 & 40 & -38.82 & 9.55 & -33.6\\ 
NGC 6218  & M 12    & 11 &4385 & 80 & -41.25 & 3.61 & -42.2\\ 
NGC 6254  & M 10    & 14 &4531 & 88 &  74.57 & 4.39 &  75.8\\
NGC 6397  &         & 13 &4798 & 79 &  22.75 & 7.10 &  18.9\\ 
NGC 6752  &         & 14 &4363 &100 & -26.96 & 6.01 & -27.9\\ 
NGC 6809  & M 55    & 14 &4440 & 77 & 174.83 & 4.33 & 174.8\\ 
NGC 6838  & M 71    & 12 &4064 & 92 & -23.97 & 2.56 & -22.8\\
NGC 7078  & M 15    & 13 &4668 & 84 &-107.34 & 5.08 &-107.0\\ 
NGC 7099  & M 30    & 10 &4600 & 84 &-188.05 & 5.20 &-181.9\\	
\hline
\end{tabular}
\label{t:snrv}
\end{table*}  

\paragraph{Star selection-} The target selection within a given cluster followed
the same criteria already illustrated for the previous GCs. We chose stars free
of any companion brighter than $V+2$ mag within a 2.5 arcsec radius, or
brighter than $V-2$ mag within 10 arcsec, with $V$ the target magnitude. Stars
near the RGB ridge line were primarily selected, while stars close to the
RGB tip were avoided to reduce problems related to model atmospheres. We used
two  fibre configurations  to maximize the number of possible target
observed  with UVES dedicated fibres (up to a maximum of seven, plus one
pointing to the sky)  while observing the bulk of programme stars with the GIRAFFE
instrument. The constraints due to mechanical limitations in the Oz-Poz fibre
positioner for FLAMES, coupled with the clusters sizes on the sky, restricted
observations to stars at some distance from the clusters centres. However, these
circumstances did not increase the risk of including field interlopers much
except for the disc cluster NGC~6171 (M~107), where we found that only five UVES
fibres targeted cluster member stars. Relevant information for all individual
target stars (cluster,  star identification, coordinates, magnitudes, and
heliocentric radial velocities) are given in Table~\ref{t:coo}
(on-line).

\begin{table*}
\centering
\caption{List and relevant information for the 202 target stars in 17 GCs.
The complete table is
available electronically only at CDS; we show here a few lines for guidance.
}
\begin{tabular}{ccccccccrr}
\hline
\hline
GC      &ID     &RA           &Dec           &$B$    &$V$    &$I$   &$K$     &RV(Hel ) &$rms$    \\
\hline
NGC~104 &  5270 & 0 24 16.877 & -72 11 49.47 & 13.654&12.272 &0.000 &  8.838 & -10.25  & 0.01  \\
NGC~104 & 12272 & 0 23 55.015 & -72 08 23.67 & 13.732&12.433 &0.000 &  9.122 & -11.96  & 0.46  \\
NGC~104 & 13795 & 0 25 13.285 & -72 07 50.43 & 13.813&12.553 &0.000 &  9.431 & -28.88  & 0.01  \\
NGC~104 & 14583 & 0 22 31.979 & -72 06 48.99 & 14.118&12.878 &0.000 &  9.831 & -11.90  & 0.02  \\
NGC~104 & 17657 & 0 25 04.269 & -72 06 39.87 & 13.697&12.252 &0.000 &  8.628 & -24.12  & 0.31  \\
\hline
\end{tabular}
\label{t:coo}
\end{table*}

\paragraph{Photometry and astrometry-} We used the available optical
photometry,  calibrated to the standard  Johnson-Cousins system, for our target
selection. The adopted photometric and astrometric data are discussed in
Carretta et al. (2006, Paper I) for NGC~2808, in Carretta et al. (2007a, Paper
II) for NGC~6752, in Carretta et al. (2007c, Paper IV) for NGC~6218 (M~12), and
in Carretta et al. (2009, Paper VII) for the remaining clusters analysed here.
The $K$ magnitudes used for the determination of effective temperature and surface
gravity (see next section) are from the 2MASS Point Source Catalogue (Skrutskie et
al. 2006). Optical and near IR magnitudes are listed in Table~\ref{t:coo}.

\paragraph{Spectroscopic data preparation-} We used the one-dimensional, 
wavelength-calibrated, reduced spectra from the ESO UVES-FLAMES pipeline (uves/2.1.1
version) as prepared by the ESO service mode personnel. Radial velocities (RVs)
were measured for each individual spectrum using many lines with
the {\sc IRAF}\footnote{IRAF is distributed by the National Optical
Astronomical Observatory, which are operated by the Association of Universities
for Research in Astronomy, under contract with the National Science Foundation }
package {\sc RVIDLINES}. Using these RVs the spectra were shifted to zero RV and
co-added for each star. The resulting heliocentric RVs and associated errors
(typically  a few hundred m s$^{-1}$) are indicated in
Table~\ref{t:coo} for  individual member stars; average values for each
programme
cluster  (with 1$\sigma$ rms scatters) are listed in Table~\ref{t:snrv} 
with the values from the Harris (1996) updated catalogue, as a comparison. In
almost all cases it was possible to disregard non member stars on the basis  of
the RVs. Only in NGC~6218 were two stars, considered probable members 
from their RVs,
rejected on the basis of the resulting abundances as field interlopers (see
also Paper IV). For the first two clusters analysed
(NGC~6218 and NGC~6752), we used the full spectral coverage of the UVES red arm
spectra, from about 4800~\AA\ to about 6800~\AA. Afterwards, we restricted the
analysis to the range 5600-6800~\AA\ in order (i) to have a complete overlap
with the spectral range of the GIRAFFE HR11 and HR13 gratings for the
intercomparison of the $EW$s, and (ii) to exploit a region in wavelength where
the $S/N$ was higher and the line crowding lower. For the three most metal-poor
clusters (NGC~6397, NGC~7078, and NGC~7099) we had to use the entire wavelength
range to increase the number of useful lines; however, line crowding was
negligible even in the blue-yellow part of their spectra.

\section{Atmospheric parameters and analysis}

\subsection{Atmospheric parameters}

Temperatures and gravities were derived for all stars using the same procedure 
described in detail in the previous papers of the series (see Papers II to 
VII)  to which we refer the reader. The only exception was NGC~2808 (Paper I).
For this  cluster, the first analysed, temperatures were obtained only from $V-K$
colours  and the Alonso et al. (1999) calibration, without using the second step
of deriving  the final adopted temperatures from a $T_{\rm eff}$ versus magnitude
relation. For  self-consistency, we used here the same temperature derivation
also for stars in  NGC~2808 observed with UVES that, furthermore, were all
observed with both instruments.Surface gravities were obtained from the position
of stars in the colour-magnitude diagram.

\begin{figure}
\centering
\includegraphics[bb=30 150 570 700, clip,scale=0.45]{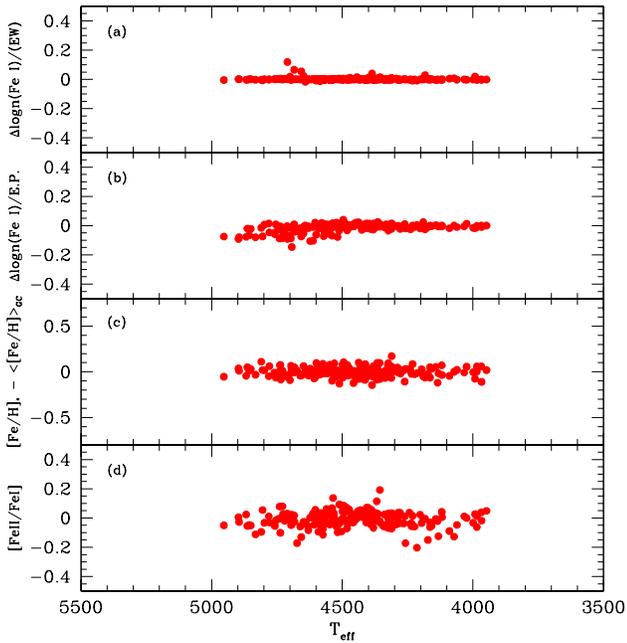} 
\caption{A graphical summary for the analysis of iron abundances in the 202 
red giant stars with UVES
spectra in the 17 GCs analysed here. $(a)$: slope of the relation 
between expected line strength and Fe~{\sc i} abundances used to
derive the $v_t$ values for each individual stars. $(b)$ slope of the relation
between Fe~{\sc i} abundances and excitation potential E.P. for each star.
$(c)$: the average [Fe/H] value for each cluster is subtracted from the
metallicity of each star in the cluster, and the differences are shown. $(d)$:
differences in iron abundances from Fe~{\sc i} and Fe~{\sc II} lines. All
quantities are plotted as a function of the effective temperature.}
\label{f:analisiFeU}
\end{figure}

Microturbulent velocities, $v_t$, were derived by eliminating trends in the 
relation between abundances from Fe neutral lines and expected line strength 
(see Magain 1984). Final metallicities were obtained by interpolating 
the model with the proper atmospheric parameters whose abundance matched  the
one derived from Fe~{\sc i} lines in  the
Kurucz (1993) grid of model atmospheres (with overshooting on).

Adopted atmospheric parameters, along with iron abundances derived from neutral
and singly ionised lines, are listed in Table~\ref{t:atmpar} (completely
available  only in electronic form at CDS) for all the 202 stars with UVES
spectra  analysed in the 17 GCs of the present work.

\begin{table*}
\centering
\caption[]{Atmospheric parameters and iron abundances. The complete table is
available only in electronic form at CDS. 
}
\begin{tabular}{rrccccrcccrccc}
\hline
GC &Star   &  $T_{\rm eff}$ & $\log$ $g$ & [A/H]  &$v_t$	     & nr & [Fe/H]{\sc i} & $rms$ & nr & [Fe/H]{\sc ii} & $rms$ \\
   &    &     (K)	&  (dex)     & (dex)  &(km s$^{-1}$) &    & (dex)	  &	  &    & (dex)         &       \\
\hline
NGC~104  &  5270 & 3999&1.01 &-0.77 &1.48 & 88 & -0.772& 0.101 & 8  & -0.804 & 0.130 \\  
NGC~104  & 12272 & 4061&1.14 &-0.75 &1.64 & 90 & -0.747& 0.102 & 6  & -0.794 & 0.134 \\  
NGC~104  & 13795 & 4183&1.30 &-0.83 &1.49 & 72 & -0.832& 0.093 & 7  & -0.836 & 0.089 \\  
NGC~104  & 14583 & 4231&1.47 &-0.68 &1.54 & 72 & -0.684& 0.070 & 7  & -0.669 & 0.098 \\  
NGC~104  & 17657 & 3992&0.99 &-0.84 &1.56 & 79 & -0.844& 0.088 & 6  & -0.818 & 0.143 \\  
 \hline
\end{tabular}
\label{t:atmpar}
\end{table*}

\subsection{Equivalent widths and iron abundances}

The adopted line lists, atomic parameters, and reference solar abundances (from
Gratton et al. 2003) are strictly  homogeneous for all stars analysed in all
clusters of the present programme.  We stress that this is the remarkable strength
of the project, aimed at obtaining  an extensive database of stellar abundances
derived in the most homogeneous way for  a significant part of the galactic GC
population.

Equivalent widths  were measured with the same automatic procedure as we
used in the analysis of GIRAFFE spectra (Papers I, II, VI, V, VII; see the
detailed description in Bragaglia et al. 2001 for the  definition of  the local
continuum around each line). Stars observed in each cluster with both 
instruments were used to register the $EW$s from GIRAFFE spectra to those 
measured here on the higher resolution UVES spectra (see also the discussion in 
Paper VII), using linear relations to correct the $EW$s. 

After this correction onto the UVES system is applied, $EW$s from 
GIRAFFE spectra agree very well with those measured even in extremely 
high-resolution spectra with quite high $S/N$ (see a lengthier discussion in Paper
II, section 2.2). 

\begin{table*}
\centering
\caption{Iron abundances from UVES spectra.}
\begin{tabular}{rrcccccl}
\hline
GC       &nr &   [Fe/H]{\sc i}              & $rms$ & $<\Delta \theta>$ & $<\Delta$(FeI)$/EW>$& $<$[Fe{\sc ii}/Fe{\sc i}]$>$ & $rms$ \\
         &stars& $\pm$stat.err.$\pm$syst.   &       &$\pm$stat.err	&$\pm$stat.err	    &	  	       &       \\
         &   &        (dex)                 &       &                   &                   &                  &       \\
\hline
NGC  104 &11 &$-$0.768$\pm$0.016$\pm$0.031 & 0.054 &$-$0.008$\pm$0.003 &   +0.003$\pm$0.004& $-$0.031& 0.073 \\
NGC  288 &10 &$-$1.305$\pm$0.017$\pm$0.071 & 0.054 &$-$0.009$\pm$0.002 & $-$0.001$\pm$0.001& $-$0.044& 0.047 \\
NGC 1904 &10 &$-$1.579$\pm$0.011$\pm$0.069 & 0.033 &  +0.004$\pm$0.002 &   +0.000$\pm$0.000&   +0.031& 0.026 \\
NGC 2808 &12 &$-$1.151$\pm$0.022$\pm$0.050 & 0.075 &$-$0.007$\pm$0.002 & $-$0.002$\pm$0.001& $-$0.015& 0.031 \\
NGC 3201 &13 &$-$1.512$\pm$0.018$\pm$0.075 & 0.065 &  +0.016$\pm$0.003 &   +0.002$\pm$0.001&   +0.049& 0.064 \\
NGC 4590 &13 &$-$2.265$\pm$0.013$\pm$0.070 & 0.047 &$-$0.069$\pm$0.010 &   +0.022$\pm$0.010& $-$0.001& 0.073 \\
NGC 5904 &14 &$-$1.340$\pm$0.014$\pm$0.064 & 0.052 &  +0.001$\pm$0.003 &   +0.001$\pm$0.002& $-$0.003& 0.034 \\
NGC 6121 &14 &$-$1.168$\pm$0.012$\pm$0.054 & 0.046 &  +0.001$\pm$0.003 &   +0.000$\pm$0.000& $-$0.002& 0.025 \\
NGC 6171 & 5 &$-$1.033$\pm$0.029$\pm$0.038 & 0.064 &  +0.009$\pm$0.006 &   +0.000$\pm$0.001& $-$0.037& 0.031 \\
NGC 6218 &11 &$-$1.330$\pm$0.013$\pm$0.066 & 0.042 &$-$0.012$\pm$0.004 &   +0.001$\pm$0.001& $-$0.037& 0.017 \\
NGC 6254 &14 &$-$1.575$\pm$0.016$\pm$0.076 & 0.059 &$-$0.017$\pm$0.004 &   +0.001$\pm$0.001& $-$0.021& 0.071 \\
NGC 6388 & 7 &$-$0.441$\pm$0.014$\pm$0.025 & 0.038 &$-$0.018$\pm$0.003 & $-$0.000$\pm$0.003&   +0.072& 0.103 \\
NGC 6397 &13 &$-$1.988$\pm$0.012$\pm$0.061 & 0.044 &$-$0.036$\pm$0.008 &   +0.000$\pm$0.001& $-$0.039& 0.019 \\
NGC 6441 & 5 &$-$0.430$\pm$0.026$\pm$0.050 & 0.058 &$-$0.028$\pm$0.013 & $-$0.002$\pm$0.001&   +0.006& 0.197 \\
NGC 6752 &14 &$-$1.555$\pm$0.014$\pm$0.074 & 0.051 &$-$0.005$\pm$0.002 &   +0.002$\pm$0.001&   +0.034& 0.017 \\
NGC 6809 &14 &$-$1.934$\pm$0.017$\pm$0.074 & 0.063 &  +0.002$\pm$0.003 &   +0.005$\pm$0.003&   +0.011& 0.036 \\
NGC 6838 &12 &$-$0.832$\pm$0.018$\pm$0.051 & 0.061 &$-$0.004$\pm$0.002 & $-$0.002$\pm$0.000& $-$0.022& 0.049 \\
NGC 7078 &13 &$-$2.320$\pm$0.016$\pm$0.069 & 0.057 &$-$0.042$\pm$0.010 & $-$0.000$\pm$0.001& $-$0.034& 0.053 \\
NGC 7099 &10 &$-$2.344$\pm$0.015$\pm$0.069 & 0.049 &$-$0.041$\pm$0.013 &   +0.000$\pm$0.000&   +0.003& 0.070 \\
\hline
\end{tabular}
\label{t:iron}
\end{table*}

Average abundances of iron derived from UVES spectra for the 17 programme 
clusters are summarised in Table~\ref{t:iron}, where we include also the 2  bulge
clusters (NGC~6388 and NGC~6441) previously analysed in Papers III and  VI.
Beside the number of stars for each cluster, we list in this table the average
value of [Fe/H] with two attached errors: the first is the statistical error and
the second represents the systematic error related to the various assumptions we
made in the analysis (see below, next section)\footnote{As discussed in Paper IV
these systematic errors only include uncertainties in the average cluster
abundances. Limitations due to e.g. the grid of model atmospheres used or to 
the reference solar abundances, which apply to all clusters, are not included.} .
The $rms$ scatter about the mean is given in the next column of
Table~\ref{t:iron}.

The average slope $\Delta \theta$ of the relation between abundances from 
Fe~{\sc i} lines and excitation potential and the average slope in the 
relation of the expected line strength versus  Fe~{\sc i} (used to derive the
microturbulent  velocity) are given in the next two columns, each with its
statistical error  attached.  Individual values of these slopes are plotted as a
function of the adopted effective temperature in the first two panels of
Fig.~\ref{f:analisiFeU} which  summarises our analysis of iron abundances in the
17 programme clusters.  Panel (a) shows how well the microturbulent velocity is
constrained by our using several tens of Fe~{\sc i}  lines and the expected line
strength: the average value is  $0.002 \pm0.001$ with $rms=0.012$ from 202
stars, stable even when a  2.5$\sigma-$clipping is applied. Panel (b) in
Fig.~\ref{f:analisiFeU} displays the slopes of the relation between abundances
from neutral Fe~{\sc i} lines and excitation potential for each analysed  star.
There is a slight trend toward more negative value for lower metallicity stars, the
average value for the 202 stars being  $-0.013 \pm0.002$ with $rms=0.029$, which
implies that we derive temperatures from colours about 45 K higher than those
we would derive from the iron excitation equilibrium. However, after excluding
32 outliers with a 2.5$\sigma-$clipping, the average slope becomes only $-0.004
\pm0.001$ (with $rms=0.013$), corresponding to an offset in temperature of only
about 14 K.

To plot the metal abundance of single stars for all clusters, we
subtracted the average metallicity of the parent cluster from each star. The
results  are plotted in panel (c) of Fig.~\ref{f:analisiFeU} and show no trend
whatsoever as a function of the temperature (on a range of $\sim 1000$ K), the
mean difference being $0.000 \pm 0.004$ with $rms=0.050$ dex, after 3 outliers
were eliminated by a 2.5$\sigma-$clipping.

Finally, the last two columns of Table~\ref{t:iron} list the average 
[Fe~{\sc ii}/Fe~{\sc i}] ratio for each cluster and its $rms$ scatter of the mean.
Individual values are plotted in the last panel (d) of Fig.~\ref{f:analisiFeU}
and the average value from 191 stars left after the usual clipping of outliers
is $-0.004 \pm 0.003$ ($rms=0.042$ dex). This difference is obviously 
negligible and it indicates that the iron ionisation equilibrium is perfectly
matched by our present abundance analysis. In turn, since surface gravities are
derived from the position in the colour-magnitude diagram, this evidence
supports the assumptions  we made in the analysis very well, namely the
temperature scale and the adopted  distance moduli and reddening values.

\begin{figure}
\centering
\includegraphics[bb=30 150 360 700, clip, scale=0.65]{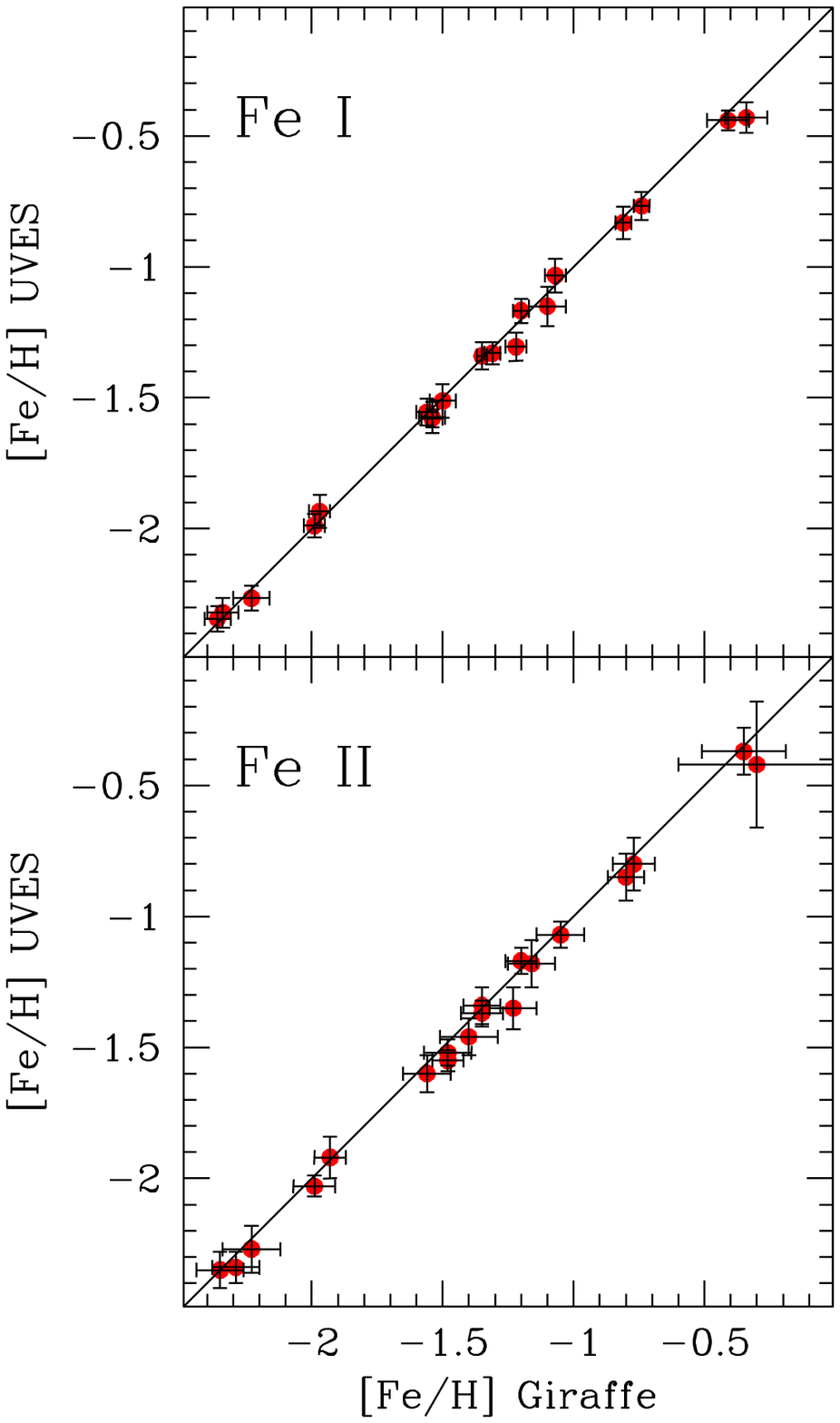}  
\caption{Comparison for [Fe/H]{\sc i} ratios (upper panel) and for [Fe/H]{\sc
ii} ratios (lower panel) obtained from GIRAFFE and UVES spectra in all the 19
programme clusters in our sample. Metal abundances from GIRAFFE are from Papers
I,II,IV,V, and VII. Metallicities from UVES are from Papers III, VI ,and the
present paper. Error bars are 1$\sigma$ $rms$ scatter and the solid line is the
line of equality.}
\label{f:FeIFeII}
\end{figure}

A comparison between iron abundances as obtained from GIRAFFE (in Papers
I, II, IV, V, and VII) and UVES spectra (Papers III, VI, and the present
paper) is plotted in Fig.~\ref{f:FeIFeII} for all  19 GCs in
our sample. In the upper panel the comparison is done for neutral iron, while
abundances from singly ionised species are used in the lower panel. All data 
points lie within 1$\sigma$ $rms$ on the line of equal abundance, convincingly 
showing how homogeneous are our final metallicities. We therefore consider 
that the samples of stars observed with both the GIRAFFE and UVES spectrographs 
can be safely merged, as done in Paper VII to increase statistics and study the 
properties of stars belonging to different stellar generations in GCs.

\section{Errors in the atmospheric parameters}

Our procedure to estimate errors is described in detail in previous papers so
will not be repeated here. In the present analysis we  followed all the steps
given in the extensive Appendix A of the companion  Paper VII. In the following
we address only differences with respect to that error analysis.

\paragraph{Sensitivities of abundance ratios to atmospheric parameters-}  We
have 173 stars with UVES spectra observed also with GIRAFFE. Since the 
temperature range is quite similar for the two samples, we adopted the
sensitivities for the  elements Fe~{\sc i}, Fe~{\sc ii}, O, and Na from Paper
VII. For the elements  not studied  in that work (i.e., Mg, Al, and Si), we
obtained the new sensitivity values by repeating our abundance  analysis for
each star and changing only one atmospheric parameter each time.  For every
cluster the adopted sensitivity to each parameter is the average of what results
from the individual stars. In doing so, we also checked  that the slope
abundance/parameters for iron were similar to those obtained  from GIRAFFE
spectra.
Table~\ref{t:sensitivityU} lists the abundance changes of Mg, Al, and Si to the
indicated variations in atmospheric parameters for the 17 clusters analysed
here.

\begin{table*}
\centering
\caption{Sensitivities of abundance ratios to errors in the atmospheric
parameters. The complete table is only available in electronic form.}
\begin{tabular}{lrrrcrrr}
\hline
& \multicolumn{3}{c}{$\Delta$T$_{\rm eff}=50$~K} &\multicolumn{1}{c}{}  &\multicolumn{3}{c}{$\Delta v_t = +0.1$~km/s} \\
	\cline{2-4} \cline{6-8} \\

cluster &$\Delta$[Mg/Fe]&$\Delta$[Al/Fe]&$\Delta$[Si/Fe]& &$\Delta$[Mg/Fe]&$\Delta$[Al/Fe]&$\Delta$[Si/Fe]\\

NGC 104 & $-$0.001&   +0.032 &$-$0.049&  & +0.019   & +0.022   & +0.026 \\
NGC 288 & $-$0.018& $-$0.012 &$-$0.052&  & +0.015   & +0.023   & +0.019 \\
\hline
& \multicolumn{3}{c}{$\Delta \log g = +0.2$~dex} &\multicolumn{1}{c}{}  &\multicolumn{3}{c}{$\Delta$ [A/H] = +0.1~dex}\\
	\cline{2-4} \cline{6-8} \\ 

cluster  &$\Delta$[Mg/Fe]&$\Delta$[Al/Fe]&$\Delta$[Si/Fe]& &$\Delta$[Mg/Fe]&$\Delta$[Al/Fe]&$\Delta$[Si/Fe] \\	       
NGC 104 & $-$0.022& $-$0.038 &  +0.018&  & $-$0.007 & $-$0.022 & +0.005 \\
NGC 288 & $-$0.009& $-$0.012 &  +0.025&  & $-$0.004 & $-$0.004 & +0.012 \\

\\	       
\hline
\end{tabular}
\label{t:sensitivityU}
\end{table*}

\paragraph{Errors in atmospheric parameters-}
Both star-to-star (internal) and systematic errors associated to the adopted
atmospheric parameters were derived as in Appendix A of Paper VII and 
are summarised in Table~\ref{t:errparU} for our 17 programme clusters. 

\begin{table*}
\centering
\caption{Star-to-star errors and cluster errors in atmospheric parameters and in the EWs from UVES}
\begin{tabular}{lrllllcrrrr}
\hline
     & \multicolumn{5}{c}{star-to-star errors}  &\multicolumn{1}{c}{} &\multicolumn{4}{c}{cluster errors} \\
	\cline{2-6} \cline{8-11}\\

cluster &$T_{\rm eff}$& $\log g$ & [A/H] & $v_t$ & EW    &   & $T_{\rm eff}$& $\log g$ & [A/H] & $v_t$\\
        &    (K)      &  dex     & (dex) & (km/s)&(dex)  &   &  (K)         &  dex     & (dex) & (km/s) \\
	&    (1)      &  (2)     & (3)   & (4)   &(5)    &   &  (6)         &  (7)     & (8)   & (9)   \\
	\cline{2-6} \cline{8-11}\\ 
NGC 104 &  	6     &  0.020   & 0.054 &  0.06 & 0.010 &   &    40        &  0.059   & 0.031 & 0.018 \\
NGC 288 &  	6     &  0.041   & 0.054 &  0.08 & 0.010 &   &    63        &  0.061   & 0.071 & 0.025 \\
NGC~1904&  	5     &  0.041   & 0.033 &  0.06 & 0.011 &   &    57        &  0.060   & 0.069 & 0.019 \\
NGC~2808&      44     &  0.020   & 0.075 &  0.06 & 0.010 &   &    42        &  0.059   & 0.050 & 0.017 \\
NGC~3201&  	4     &  0.041   & 0.065 &  0.05 & 0.011 &   &    62        &  0.061   & 0.075 & 0.014 \\
NGC~4590&  	4     &  0.041   & 0.047 &  0.38 & 0.020 &   &    69        &  0.061   & 0.070 & 0.105 \\
NGC~5904&      12     &  0.041   & 0.052 &  0.05 & 0.012 &   &    54        &  0.060   & 0.064 & 0.013 \\
NGC~6121&  	4     &  0.041   & 0.046 &  0.04 & 0.011 &   &    54        &  0.060   & 0.054 & 0.011 \\
NGC~6171&  	2     &  0.041   & 0.064 &  0.07 & 0.014 &   &    26        &  0.057   & 0.038 & 0.031 \\
NGC~6218&  	6     &  0.020   & 0.042 &  0.04 & 0.011 &   &    48        &  0.059   & 0.066 & 0.012 \\
NGC~6254&  	4     &  0.041   & 0.059 &  0.09 & 0.012 &   &    67        &  0.061   & 0.076 & 0.024 \\
NGC~6397&  	4     &  0.041   & 0.044 &  0.08 & 0.016 &   &    64        &  0.060   & 0.061 & 0.022 \\
NGC~6752&  	5     &  0.041   & 0.051 &  0.03 & 0.011 &   &    58        &  0.060   & 0.074 & 0.008 \\
NGC~6809&  	5     &  0.041   & 0.063 &  0.13 & 0.013 &   &    58        &  0.060   & 0.074 & 0.035 \\
NGC~6838&  	5     &  0.041   & 0.061 &  0.06 & 0.012 &   &    45        &  0.059   & 0.051 & 0.017 \\
NGC~7078&  	5     &  0.041   & 0.057 &  0.26 & 0.020 &   &    67        &  0.061   & 0.069 & 0.072 \\
NGC~7099&  	5     &  0.041   & 0.049 &  0.12 & 0.018 &   &    71        &  0.061   & 0.069 & 0.038 \\

\\
\hline
\end{tabular}
\label{t:errparU}
\end{table*}

The main differences (in particular w.r.t. the estimates from GIRAFFE spectra) 
are:
\begin{itemize}
\item In the case of NGC~2808 the adopted temperatures were not derived from the
relation with the magnitude; as a result, the internal (star-to-star) error in
effective temperature turns out to be about 8 times more than in the other
clusters. This is good proof of the soundness of our approach.
\item For the other clusters, errors in $T_{\rm eff}$ and gravity are as
previously estimated in Paper VII
\item Errors in metallicity do change because they are estimated from the $rms$
scatter in [Fe/H] of all analysed stars.
\item Errors in $v_t$ are generally less than those from GIRAFFE spectra due to
the much larger number of Fe~{\sc i} lines used to estimate the microturbulent
velocity
\item Errors from uncertainties in the measurement of $EW$s are  lower, since
they were obtained from the $rms$ in Fe~{\sc i} divided by the square root of the
typical number of measured lines, usually higher in UVES spectra, in particular
in the case of iron.
\end{itemize}

Finally, in Table~\ref{t:errabuU} we show the errors in the abundances of Fe
(from neutral and singly ionised species), O, Na, Mg, Al, and Si as due to
uncertainties in each atmospheric parameter and to errors in the $EW$
measurement. The sums in quadrature of the three major 
($T_{\rm eff}$, $v_t$, $EW$) or of all error sources (last two columns in 
Table~\ref{t:errabuU}) represent the expected typical star-to-star error in each
abundance ratio.

\begin{table*}
\caption{Error in element ratios due to star-to-star errors in atmospheric
parameters and in the EWs for UVES.The complete table is only available in 
electronic form. }
\scriptsize \centering
\begin{tabular}{lllllrlcccl}
\hline
& \multicolumn{6}{c}{errors in abundances due to:} &\multicolumn{1}{c}{} &\multicolumn{2}{c}{total star-to-star error} &\\
  \cline{2-7} \cline{9-10}\\
  
            & $T_{\rm eff}$& $\log g$ & [A/H]   & $v_t$  &$<$nr$>$ & EW  &  & $T_{\rm eff}$+$v_t$+EW &     all &\\
$[$Fe/H$]$I &     +0.004   &  +0.001  &  +0.003 &$-$0.021& 81   & 0.010  &  &  	 0.024  	     &   0.024 &NGC 104\\
$[$Fe/H$]$II&   $-$0.006   &  +0.004  &  +0.011 &$-$0.009&  7   & 0.035  &  &  	 0.037  	     &   0.038 &\\
$[$O/Fe$]$  &   $-$0.003   &  +0.003  &  +0.009 &  +0.021&  2   & 0.065  &  &  	 0.068  	     &   0.069 &\\
$[$Na/Fe$]$ &     +0.001   &$-$0.002  &  +0.000 &  +0.005&  4   & 0.046  &  &  	 0.046  	     &   0.046 &\\
$[$Mg/Fe$]$ &     +0.000   &$-$0.002  &$-$0.004 &  +0.011&  3   & 0.053  &  &  	 0.054  	     &   0.054 &\\
$[$Al/Fe$]$ &     +0.004   &$-$0.004  &$-$0.012 &  +0.013&  2   & 0.065  &  &  	 0.066  	     &   0.068 &\\
$[$Si/Fe$]$  & $-$0.006 & +0.002 & +0.003 & +0.016 & 8 &  0.033 & & 0.037 &  0.037  &\\
\\
\hline
\end{tabular}
\label{t:errabuU}
\end{table*}

The observed scatter in Fe abundances will be compared to errors in a
forthcoming paper. We only note here that as far as iron is concerned, this
scatter is generally very small, so that the GCs can  be still
considered as monometallic aggregates. In turn, the implication  is that
products of supernovae nucleosynthesis must be very well mixed  in the gas from
which any protocluster started to form its stars. This  is an inescapable
constraint and requisite for any theoretical model of  GCr
formation.

Summarising, typical star-to-star errors for the 17 clusters in the present 
work are on average 0.088 dex in [O/Fe], 0.061 dex in [Na/Fe], 0.070 dex in
[Mg/Fe], 0.078 dex in [Al/Fe], and 0.049 dex in [Si/Fe].

\section{Results: proton-capture elements}

In this section we  discuss abundances for the light elements O, Na, Mg, Al,
and Si participating in the chains of proton-capture reactions in H-burning at
high temperature. These have been recognised for a long time as those
responsible for generating the pattern of chemical composition typical of (and
restricted only  to) GC stars (see e.g. Denisenkov and Denisenkova
1989; Langer et al. 1993; and Gratton et al. 2004 for a recent review).

Abundances for these elements were obtained in the 202 RGB stars of the sample
of 17 clusters analysed here from measured $EW$s.

Oxygen abundances are obtained from the forbidden  [O {\sc i}] lines at 6300.3 
and 6363.8~\AA. We used a synthetic spectrum template (as described in Paper I) 
to clean each individual star spectra from telluric contamination by H$_2$O 
and O$_2$ lines in the region around the oxygen line at 6300.3~\AA. The 
spectra were then shifted to zero RV and coadded. 
On the basis of previous experience, we consider as  
negligible the contributions to the measured oxygen $EW$ due to the high
excitation Ni {\sc i} line at  6300.34~\AA\ and CO formation in our
programme 
stars. Furthermore, as discussed in Papers III and VI, even when present these
effects  have opposite  signs and compensate each other.

The Na abundances, derived from the doublets at 5682-88~\AA\ and 6154-60~\AA, 
were corrected for effects of departures from the LTE assumption using the
prescriptions by Gratton et al. (1999). Mg abundances are typically based on two
to three high excitation lines (Mg~{\sc i} 5711.09, 6318.71 and 6319.24~\AA) and Al 
abundances are all derived from the subordinate doublet at 6696-98~\AA, the
only  one lying in the spectral region covered by UVES red arm spectra. 
The Si
abundances are presented here because we found that in some clusters this element
is partly involved, actually produced, in proton-capture reactions (see
below)\footnote{The remaining $\alpha-$elements for the combined large sample of
stars with GIRAFFE and UVES spectra will be discussed in a forthcoming paper
devoted to the scale of $\alpha-$ and Fe-group elements in GCs.}. 
The Si abundances
are from $EW$s of several transitions in the spectral range 5645-6145~\AA;
atomic parameters are from Gratton et al. (2003). 

\begin{table*}
\centering
\caption[]{Abundances of proton-capture elements in the 202 stars analysed. The
complete table is available only electronically at CDS.}
\begin{tabular}{rrccrccrccrccrcc}
\hline
Star        & n & [O/Fe] & rms & n & [Na/Fe] & rms & n & [Mg/Fe] & rms & n & [Al/Fe] & rms &n & [Si/Fe] & rms\\ 
            &   & (dex)  &     &   &  (dex)  &     &   & (dex)   &     &   &  (dex)  &     &\\   
\hline
NGC 104 05270 & 2 &  +0.122 &0.059 & 4 & +0.692 &0.063 & 3 & +0.550 &0.133 &  2 & +0.542 &0.091 & 8 &+0.429 &0.065 \\ 
NGC 104 12272 & 2 &  +0.270 &0.025 & 4 & +0.603 &0.068 & 3 & +0.573 &0.065 &  2 & +0.535 &0.048 &10 &+0.429 &0.100 \\ 
NGC 104 13795 & 2 &  +0.359 &0.028 & 4 & +0.301 &0.078 & 3 & +0.502 &0.140 &  2 & +0.285 &0.055 & 8 &+0.384 &0.086 \\ 
NGC 104 14583 & 2 &  +0.344 &0.035 & 4 & +0.433 &0.070 & 3 & +0.545 &0.055 &  2 & +0.436 &0.053 & 9 &+0.417 &0.055 \\ 
NGC 104 17657 & 2 &$-$0.068 &0.136 & 4 & +0.852 &0.082 & 3 & +0.496 &0.161 &  2 & +0.958 &0.141 & 9 &+0.364 &0.161 \\ 
\hline
\end{tabular}
\label{t:protonU}
\end{table*}

In Table~\ref{t:protonU} we list the abundances for the elements involved in
proton-capture reactions at high temperature, derived for all stars with UVES
spectra in the 17 programme clusters (the complete table is available only in 
electronic form at CDS). The table gives for each star the identification, the
number of lines used, the abundance ratios, and the rms values; for O and Al, we
distinguish between actual detections and upper limits.

\begin{table*}
\centering
\caption[]{Mean abundance ratios in all the 19 GCs
of our project.}
\begin{tabular}{rrccrccrccrccrcc}
\hline
Star        & n & [O/Fe] & rms & n & [Na/Fe] & rms & n & [Mg/Fe] & rms & n & [Al/Fe] & rms  & n & [Si/Fe] & rms  \\ 
            &   & (dex)  &     &   &  (dex)  &     &   & (dex)   &     &   &  (dex)  &      &	&  (dex)  &	 \\   
\hline
NGC ~104 & 11 &  +0.25 &0.15 & 11 & +0.53   &0.15 & 11 &  +0.52  &0.03 & 11 &  +0.52  &0.17 & 11& +0.40 & 0.02 \\
NGC ~288 & 10 &  +0.34 &0.14 & 10 & +0.29   &0.28 & 10 &  +0.45  &0.03 & 10 &  +0.40  &0.09 & 10& +0.37 & 0.03 \\
NGC 1904 &  9 &  +0.05 &0.29 & 10 & +0.42   &0.29 & 10 &  +0.28  &0.06 &  8 &  +0.64  &0.41 & 10& +0.29 & 0.03 \\
NGC 2808 & 12 &  +0.07 &0.36 & 12 & +0.20   &0.24 & 12 &  +0.20  &0.25 & 12 &  +0.40  &0.49 & 12& +0.28 & 0.05 \\
NGC 3201 & 13 &  +0.03 &0.28 & 13 & +0.16   &0.31 & 13 &  +0.34  &0.04 & 13 &  +0.14  &0.38 & 13& +0.30 & 0.05 \\
NGC 4590 & 13 &  +0.41 &0.11 & 12 & +0.33   &0.19 & 13 &  +0.35  &0.06 & 13 &  +0.74  &0.18 & 13& +0.40 & 0.05 \\
NGC 5904 & 14 &  +0.08 &0.23 & 14 & +0.25   &0.24 & 14 &  +0.41  &0.07 & 14 &  +0.27  &0.29 & 14& +0.30 & 0.05 \\
NGC 6121 & 14 &  +0.26 &0.10 & 14 & +0.40   &0.15 & 14 &  +0.55  &0.03 & 14 &  +0.60  &0.05 & 14& +0.52 & 0.06 \\
NGC 6171 &  5 &  +0.12 &0.13 &  5 & +0.49   &0.15 &  5 &  +0.51  &0.04 &  5 &  +0.39  &0.07 &  5& +0.53 & 0.08 \\
NGC 6218 & 11 &  +0.34 &0.14 & 11 & +0.30   &0.27 & 11 &  +0.52  &0.04 & 11 &  +0.20  &0.18 & 11& +0.35 & 0.06 \\
NGC 6254 & 14 &  +0.41 &0.15 & 14 & +0.17   &0.19 & 14 &  +0.49  &0.04 & 10 &  +0.41  &0.37 & 14& +0.28 & 0.05 \\
NGC 6388 &  7 &$-$0.30 &0.16 &  7 & +0.59   &0.16 &  7 &  +0.21  &0.07 &  7 &  +0.69  &0.24 &  7& +0.32 & 0.10 \\
NGC 6397 & 13 &  +0.29 &0.09 & 13 & +0.18   &0.19 & 13 &  +0.46  &0.04 &    &         &     & 13& +0.34 & 0.05 \\
NGC 6441 &  5 &  +0.12 &0.20 &  5 & +0.55   &0.15 &  5 &  +0.34  &0.09 &  5 &  +0.30  &0.15 &  5& +0.33 & 0.11 \\
NGC 6752 & 14 &  +0.16 &0.28 & 14 & +0.33   &0.27 & 14 &  +0.50  &0.05 & 14 &  +0.41  &0.33 & 14& +0.38 & 0.05 \\
NGC 6809 & 14 &  +0.16 &0.11 & 14 & +0.38   &0.23 & 14 &  +0.47  &0.10 & 14 &  +0.49  &0.32 & 14& +0.38 & 0.06 \\
NGC 6838 & 12 &  +0.31 &0.13 & 12 & +0.45   &0.16 & 12 &  +0.49  &0.04 & 12 &  +0.50  &0.15 & 12& +0.38 & 0.06 \\
NGC 7078 & 13 &  +0.34 &0.19 & 13 & +0.20   &0.25 & 13 &  +0.45  &0.19 & 13 &  +0.57  &0.26 & 10& +0.43 & 0.10 \\
NGC 7099 & 10 &  +0.46 &0.20 & 10 & +0.35   &0.25 & 10 &  +0.51  &0.04 & 10 &  +0.77  &0.32 &  9& +0.34 & 0.07 \\
\hline
\end{tabular}
\label{t:meantabU}
\end{table*}

We list in Table~\ref{t:meantabU} the average abundances of  O, Na, Mg, Al,  and
Si, together with the number of stars in each cluster and the 1$\sigma$ rms
scatter of the mean. In this table we include for completeness the ratios found
for the two bulge clusters previously analysed in Papers III and VI. A simple
comparison with star-to-star errors of  Table~\ref{t:errabuU} shows that the
observed spreads in these element ratios  clearly exceed those expected from
uncertainties in the analysis, apart from  the [Mg/Fe] and the [Si/Fe]
ratios in some clusters.

\begin{figure*}
\centering
\includegraphics[scale=0.8]{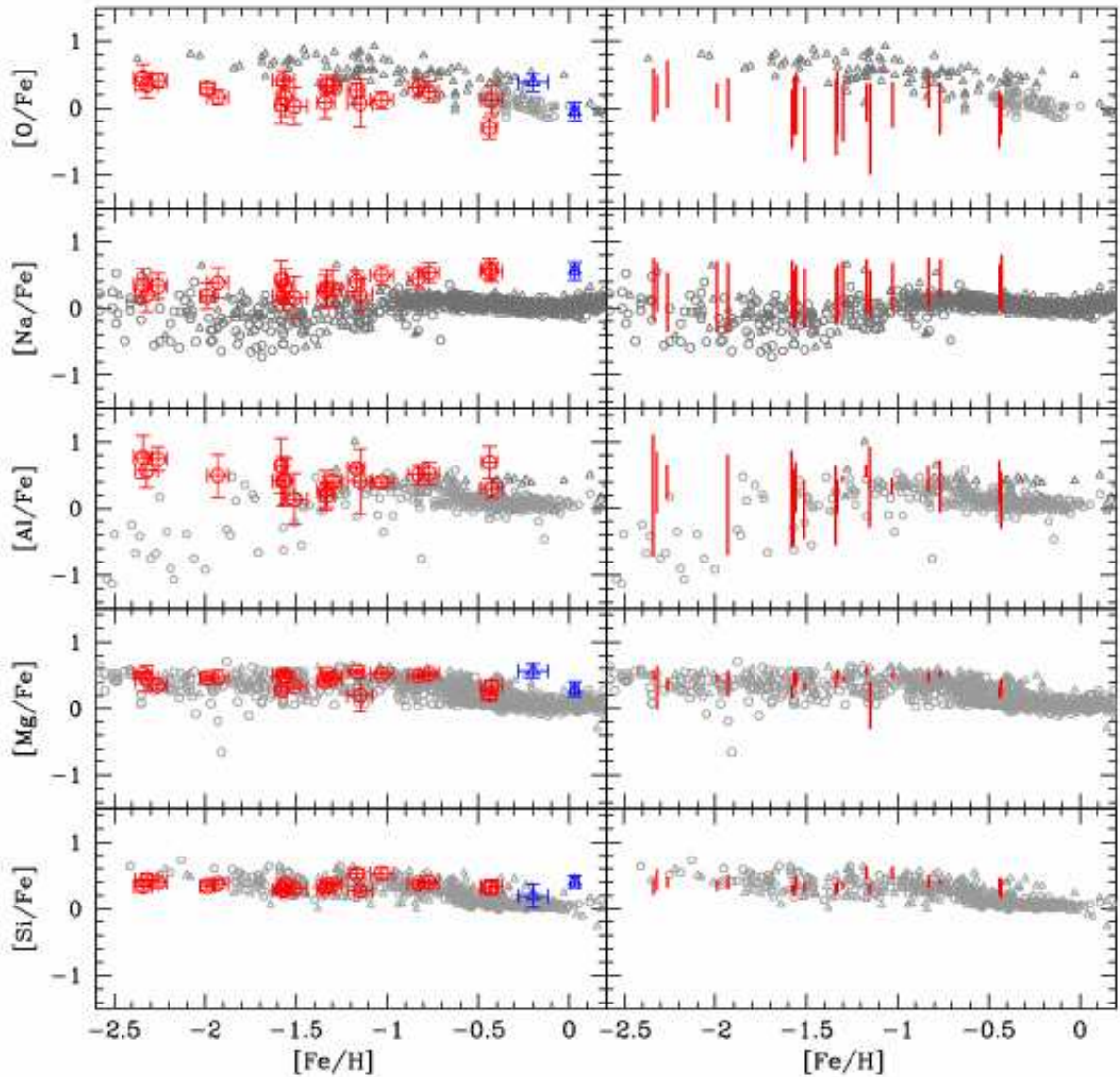} 
\caption{Left panels: average ratios of O, Na, Al, Mg, and Si  found in 
all 19 clusters of our project (large open circles), as a function of 
the metallicity, compared to the same ratios in field stars.   We added the two
bulge clusters NGC~6528 and NGC~6553,  indicated by (blue) open triangles, also
analysed by our group (Cohen et al.  1999; Carretta et al. 2001), as an
extension to higher metallicities. From  top to bottom: [O/Fe] average ratios
superimposed on values from field stars  from Gratton et al. (2003; empty
squares), Fulbright et al. (2007; open  triangles), and Reddy et al. (2003; open
pentagons); average  [Na/Fe] ratios  for our clusters with ratios for field
stars from Gratton et al. (2003),  Fulbright et al. (2007), and the compilation
by Venn et al. (2004; open circles); [Al/Fe] ratios are compared to values for
field stars from Fulbright et al. (2007;  open triangles), Fulbright (2000),
Johnson (2002), Reddy et al. (2003), Jonsell  et al. (2005) and Gehren et al.
(2006): the last five studies are indicated by  empty pentagons, collectively;
finally, in the bottom panel average [Mg/Fe]   ratios are compared to values for
field stars from Gratton et al. (2003),  Fulbright et al. (2007), and Venn et
al. (2004), plotted as empty squares,  triangles and circles, respectively.
Error bars for the cluster averages are  1$\sigma$ rms scatters of the mean.
Right panels: we show the excursions for  O, Na, Al, Mg, and Si in the 19 GCs of
our sample, also  superimposed on the same field stars.}
\label{f:gclight}
\end{figure*}

This is immediately evident also from Fig.~\ref{f:gclight} (left panels), where
for each cluster (including NGC~6388 and NGC~6441) the average ratios of the
light  elements [O/Fe], [Na/Fe], [Mg/Fe], [Al/Fe], and [Si/Fe] involved in the
proton-capture reactions are superimposed on the same ratios for a large number
of field stars spanning the same range in metallicity. Element ratios from field
stars are from literature studies, mainly from the collection by Venn et al. 
(2004), and from abundance analyses not used by the Venn et al. compilation:  
Gratton et al. (2003), Fulbright et al. (2007), Reddy et al. (2003), Fulbright
(2000), Johnson (2002), Jonsell et al. (2005), Gehren et al. (2006). The rms
scatters associated in each study with abundances of field stars are typically
less than a few hundredths of a dex, much smaller than those derived for the
average abundances found in GCs for these elements. 

From Table~\ref{t:meantabU} and Fig.~\ref{f:gclight}, it is possible to appreciate
how cluster stars differ from their analogues in the galactic fields
(irrespective of their belonging to the halo, disc, and bulge components). At
every metallicity, the oxygen abundances are depleted on average and those of Na
and Al are enhanced  in GCs, with respect to the same ratios in field stars. The
abundances of all these elements show large star-to-star variations that are not
an effect of the analysis, but reflect a true change in chemical composition in
a large fraction of stars (see also Paper VII). Differences between GCs and
field populations are much less for Mg and Si, which also show much less
spread. The smaller scatter for Mg and Si with respect to Al is in part an
effect of their much larger primordial abundance; however, it also signals that
 only a fraction of
the original Mg is transformed into Al and Si in the case of the Mg-Al cycle,
at  variance with the case of the Ne-Na cycle, where almost all $^{22}$Ne is 
transformed into $^{23}$Na. In turn, this is partly because of
the higher temperature required for the reactions to occur in the two cycles 
(25 MK and 70 MK, respectively), and of complexity of the Mg-Al
cycle itself (some of the original $^{24}$Mg is transformed into $^{25}$Mg and
$^{26}$Mg, rather than in $^{27}$Al: see Yong et al. 2003). We come back to
this point in next section.

On the other hand, if we consider the envelopes of the abundance distributions
(upper for O and Mg; lower for Na, Al, and Si), we find good agreement between
the abundances for the GCs and those for the field stars. This suggests that the
abundances of the primordial population in GCs are close to those of the field 
stars -see also the right panels of  Fig.~\ref{f:gclight}, where the entire
range of abundances in each cluster is shown.  On the whole, these are not new
results (see e.g. Gratton et al. 2000 for  a discussion of abundance variations
in light elements in field stars as opposed  to cluster stars), but in our study
this comparison is based for the first time  on a sample of cluster stars
comparable in size with the samples of field stars. There are, however,
subtler points that should be discussed, such as the large spread of the minimum values  of [Al/Fe] ratios
in the various clusters. We will  re-examine this
point in more detail in a forthcoming paper.

\subsection{The Na-O anticorrelation}

\begin{figure*}
\centering
\includegraphics[bb=10 160 570 600, clip, scale=0.89]{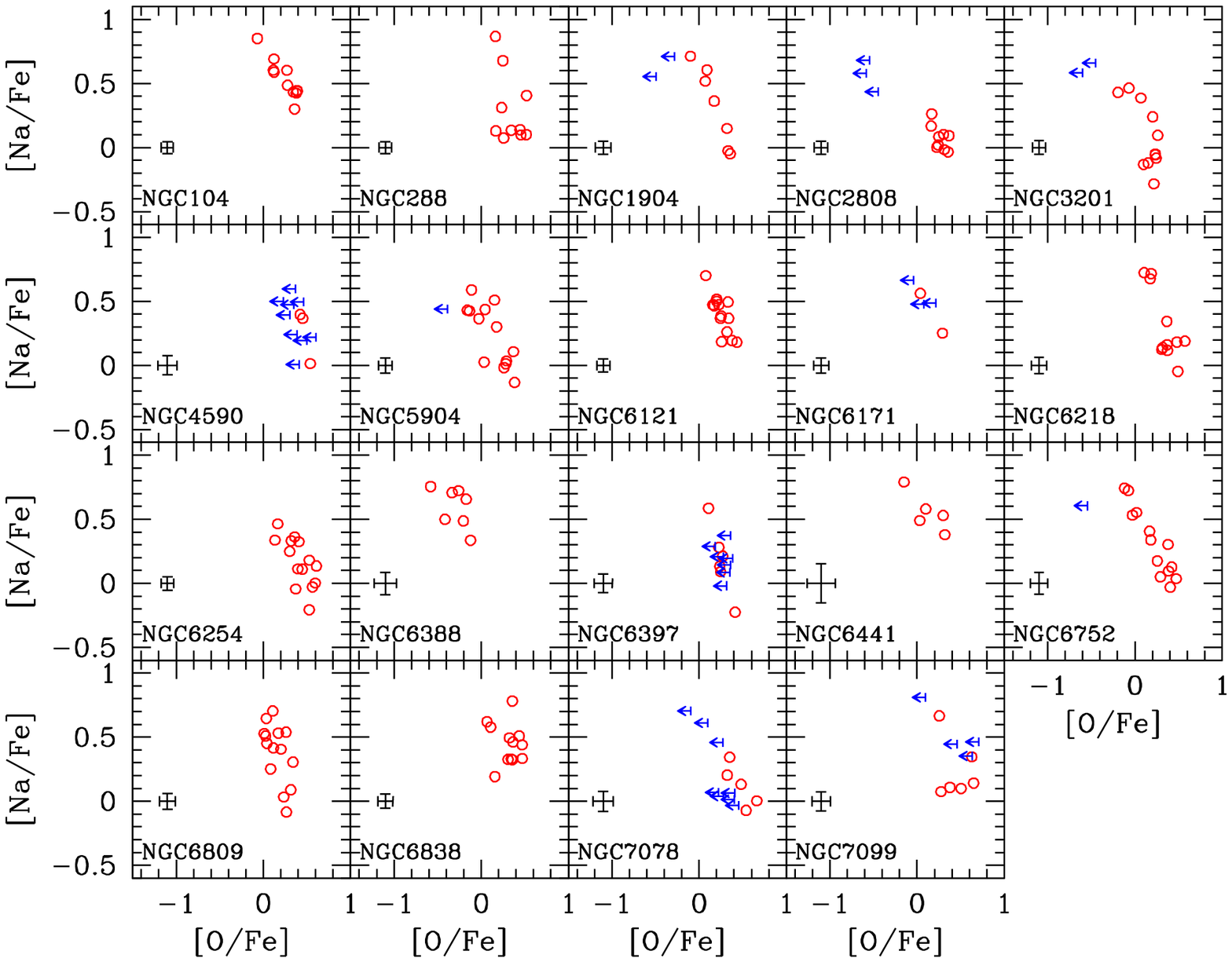} 
\caption{The Na-O anticorrelation from UVES spectra observed in all the 19 
GCs of our project (including NGC~6388 from Paper VI and NGC~6441
from Paper III). Star-to-star error bars (see Sect.4) are indicated
in each panel. Upper limits in O abundances are shown as arrows, detections are
indicated as open circles.}
\label{f:tutteantiu}
\end{figure*}

The present UVES spectra clearly reveal the classical anticorrelations between 
proton-capture elements, in spite of the limited number of UVES fibres
available  in each pointing that hampers a statistically meaningful study like
the one  on GIRAFFE data. In Fig.~\ref{f:tutteantiu} we show the Na-O
anticorrelations we  obtain using only UVES observations for the 19 clusters
observed within our programme (data from this paper, complemented by those for
NGC~6441 from Paper  III and NGC~6388 from Paper VI). Expected star-to-star
error bars are plotted  in each panel. Although limited in number in each
cluster, these additional data are very useful in some cases: as explained in the
companion Paper VII, the higher resolution and quality of UVES spectra are
crucial where only a few upper limits in O abundances were extracted from a
total of more than a hundred stars observed with GIRAFFE spectra, as happened
for NGC~6397. In fact, the extreme  homogeneity of our whole procedure allows
the two sample of stars with UVES and  GIRAFFE spectra to be safely merged, with
gain in statistics. In Paper VII, whenever  possible we replaced/complemented O
and Na abundances obtained from GIRAFFE spectra  with those from the UVES ones
when studying the different stellar populations in GCs.

\begin{figure*}
\centering
\includegraphics[scale=0.30]{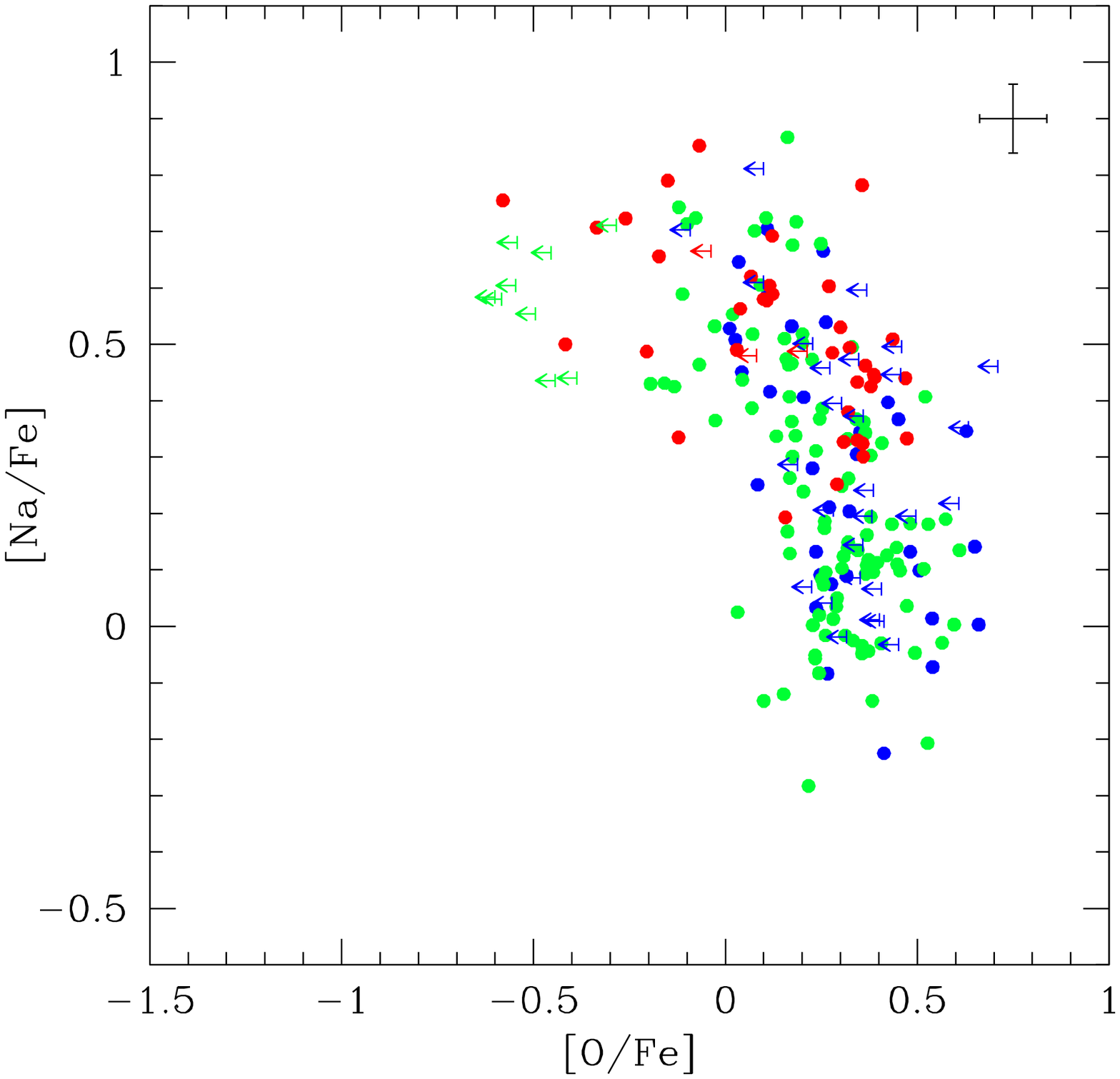}\includegraphics[scale=0.30]{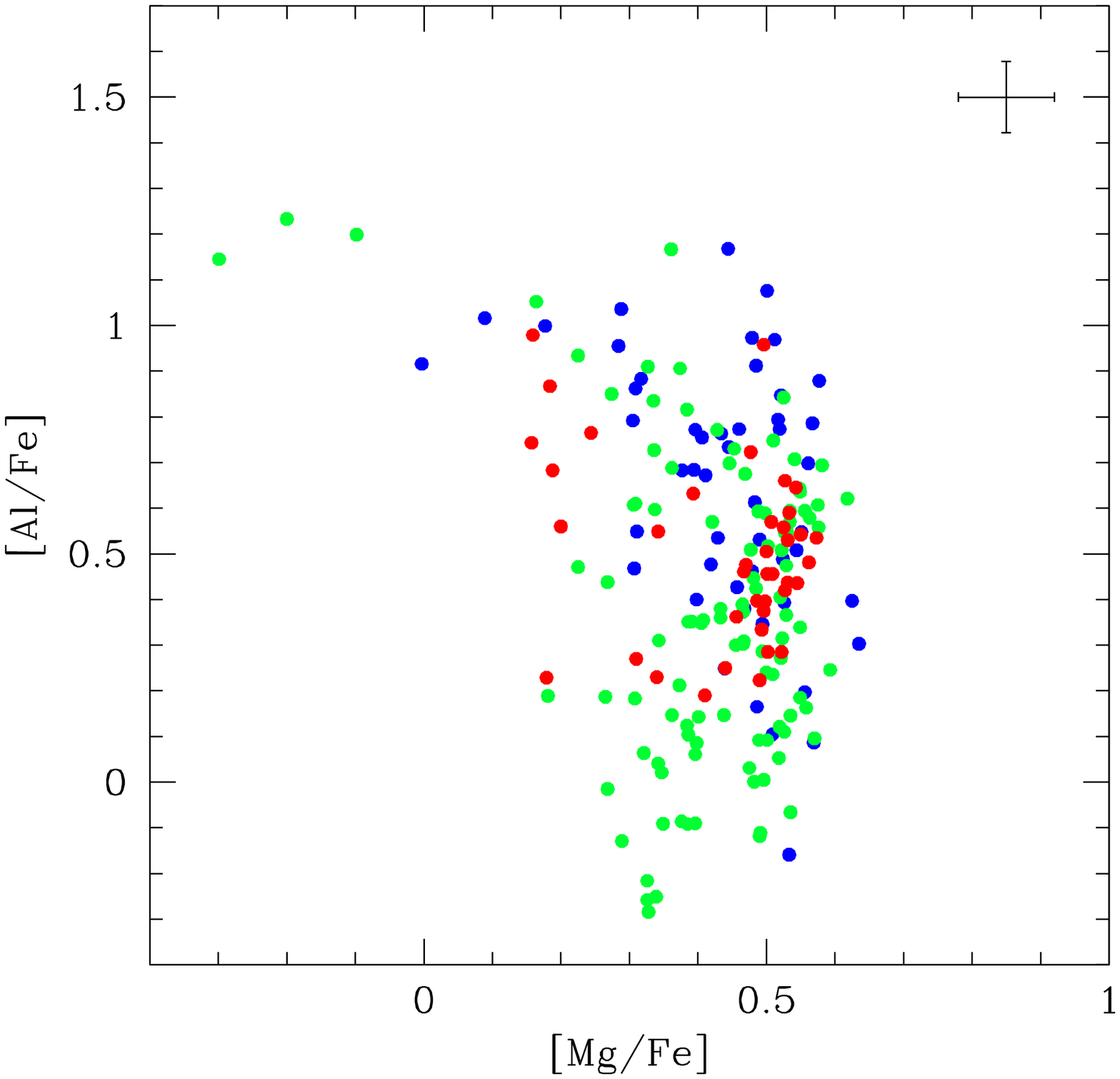}\includegraphics[scale=0.30]{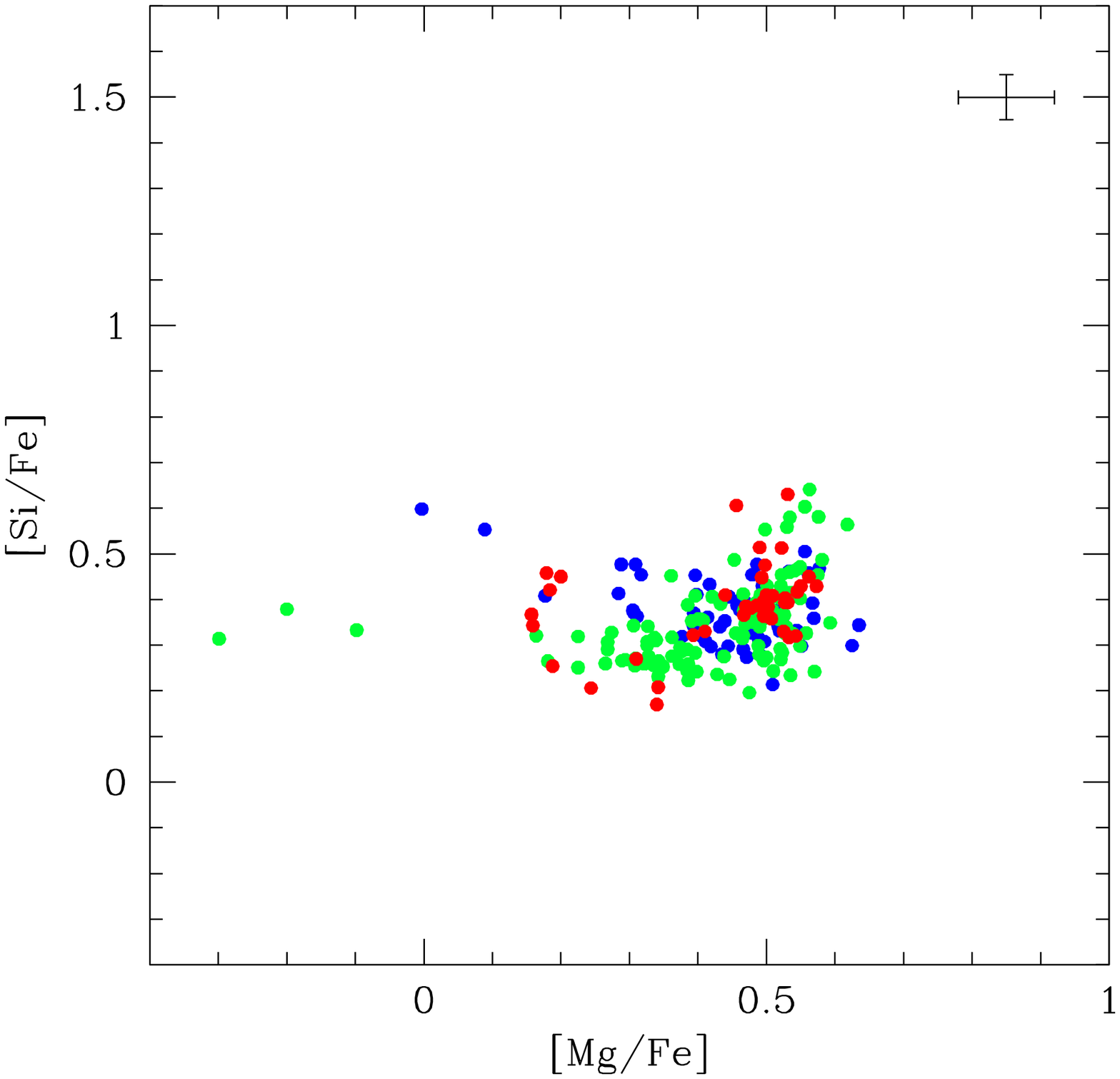}
\caption{Left panel: [Na/Fe] ratios as a function of [O/Fe] ratios in the 214
red giants with UVES spectra from the 19 clusters analysed (including NGC~6441 and
NGC~6388). Stars are colour-coded according to cluster metallicity:  in red
stars of metal-rich clusters  ($-1.1 < $[Fe/H] $< -0.4$ dex), in green giants of
clusters of intermediate metallicity ($-1.8 < $[Fe/H] $< -1.1$ dex) and blue
colours for stars in the most metal-poor clusters ($-2.4 < $[Fe/H] $< -1.8$
dex). Arrows indicate upper limits in O abundances. Central panel: the same, and
with the same colour-coding. for the [Mg/Fe] ratios as a function of [Al/Fe]
ratios. Right panel: the same, but for [Mg/Fe] and [Si/Fe].}
\label{f:anti}
\end{figure*}

The global Na-O anticorrelation from UVES spectra for all programme clusters  is
plotted in the left panel of Fig.~\ref{f:anti}. Average error bars are  also
shown in this panel. The stars are colour-coded according to arbitrarily 
defined metallicity bins: stars of metal-rich clusters ($-1.1  <
$[Fe/H] $< -0.4$ dex), giants of clusters of intermediate metallicity 
($-1.8 < $[Fe/H] $< -1.1$ dex), and finally  stars in 
the most metal-poor clusters ($-2.4 < $[Fe/H] $< -1.8$ dex).
In the Na-O plane we can see that metal-rich clusters seem to start at the same
average high oxygen values of the other GCs, but from a higher Na value, at the
O-rich, Na-poor extreme of the anticorrelation. However, this is not surprising,
because it is  assessed that the plateau of minimum Na established by
supernovae nucleosynthesis is a function of metallicity. It is higher for
[Fe/H]$>-1$ dex, as is also evident from the literature data of field stars
shown in Fig.~\ref{f:gclight}. The largest extension of the Na-O anticorrelation
seems to be reached in intermediate-metallicity clusters. Unpolluted stars (likely
first-generation stars) in the most metal-poor clusters populate the same locus
of unpolluted stars of GCs with moderate metal abundances; however, when the
cluster metallicity is  very low, the stars do not seem to show very large
O-depletion. From this sample alone, it is not clear whether (and how much) this
result might be affected by the difficulty of observing weak O lines in
metal-poor, rather warm giants.

From the left panel of Fig.~\ref{f:anti} we can see that the low-O 
([O/Fe]$\lsim -0.4$ dex) tail of the anticorrelation almost entirely
come from stars
measured in two clusters, NGC~2808 (three stars) and NGC~3201 (two
stars)\footnote{The other two stars below [O/Fe]$=-0.4$ dex are from NGC~5904
and NGC~1904, respectively}.  Actually, all $EW$s measured in these stars are
upper limits and, from the quality of the spectra, we cannot determine at the moment
what is the real O abundance. We note, however, that these three
stars in NGC~2808 also have a subsolar [Mg/Fe] ratio (they actually  define the
low-Mg tail in the Mg-Al anticorrelation shown in the middle panel of
Fig.~\ref{f:anti}), hence they are all good candidates for having chemical
compositions severely altered by nuclear reactions involving all the
proton-capture chains. In contrast, the other O-poor stars in this region
from NGC~1904, NGC~3201, and NGC~5904 do not show any strong depletion of Mg, 
falling not much below the average value of the respective clusters.  Finally,
the last two stars in the O-poor tail belong to the metal-rich and quite
massive cluster NGC~6388 (see Paper VI); also for these stars, the Mg abundances
are not exceptionally low, since very close to the cluster average.

\subsection{The Mg-Al cycle}

\begin{figure*}
\centering
\includegraphics[bb=10 160 570 600, clip, scale=0.89]{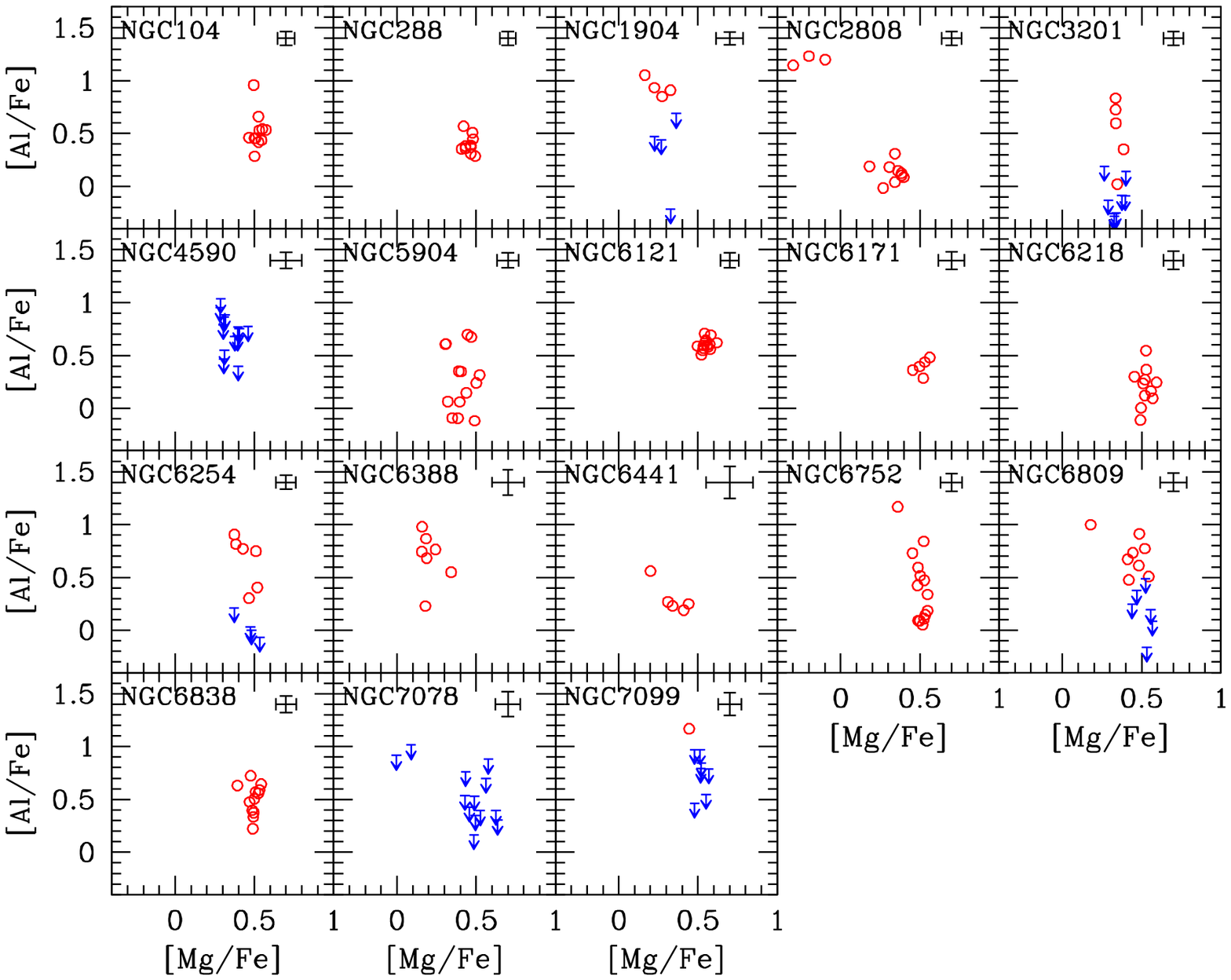} 
\caption{The Mg-Al anticorrelation from UVES spectra observed in 18 of the 19 
GCs of our project (including NGC~6388 from Paper VI and NGC~6441
from Paper III, but excluding NGC~6397, for which we did not measure Al). 
Star-to-star error bars (see Sect.4) are indicated in each panel. Upper limits
are shown as arrows, detections as open circles.}
\label{f:tutteantiu2}
\end{figure*}

\begin{figure*}
\centering
\includegraphics[bb=10 160 570 600, clip, scale=0.89]{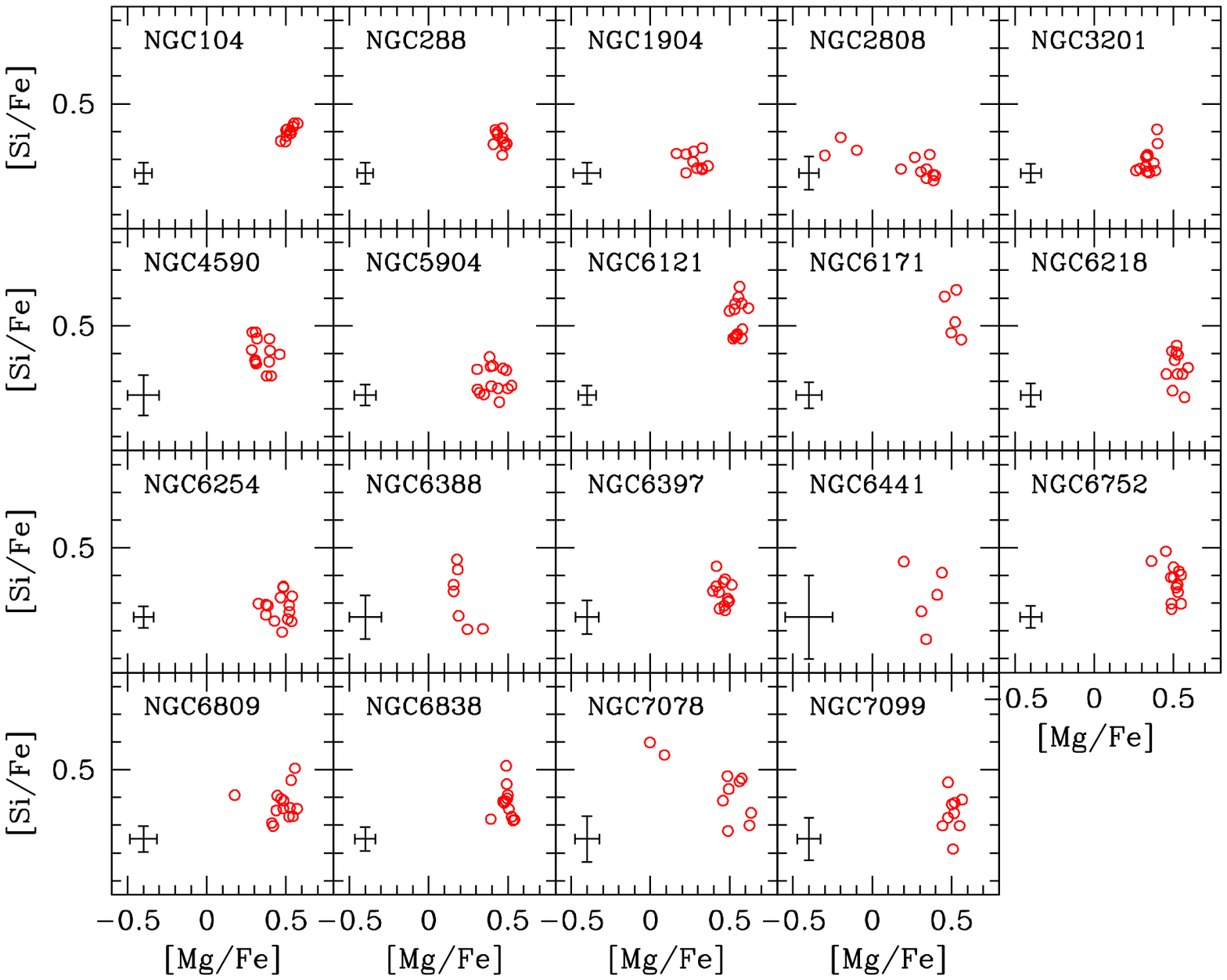} 
\caption{The Mg-Si relation from UVES spectra of the 19  GCs of
our project (including NGC~6388 from Paper VI and NGC~6441 from Paper III). 
Star-to-star error bars (see Sect.4) are indicated in each panel. }
\label{f:tutteantiu3}
\end{figure*}

Let us start by considering the abundances of the three elements involved in
the  Mg-Al cycle for which we have data (Mg, Al, and Si). We might expect some
degree of anticorrelation between Mg and Al, and possibly even between Mg and
Si, if the burning temperature is high enough.  The  central and right panels of
Fig.~\ref{f:anti} show the  global anti-correlations that we derive from our
sample of giants with UVES spectra.  There is a considerable scatter in these
plots. The Mg-Al run presents the same  features already found by previous
studies (see e.g. Cohen and Melendez 2005; Sneden et al. 2004). The star-to-star
variations in Mg abundances are limited to a range of about 0.5 dex, in most
cases much less, apart from very few exceptions; we stress  that all the [Mg/Fe]
values $\lsim 0.2$ dex were checked by eye inspection of the  spectral tracings.
In the case of Mg and Si, there is no obvious anti-correlation:  rather, there
is a global positive correlation.

\begin{figure}
\centering
\includegraphics[scale=0.45]{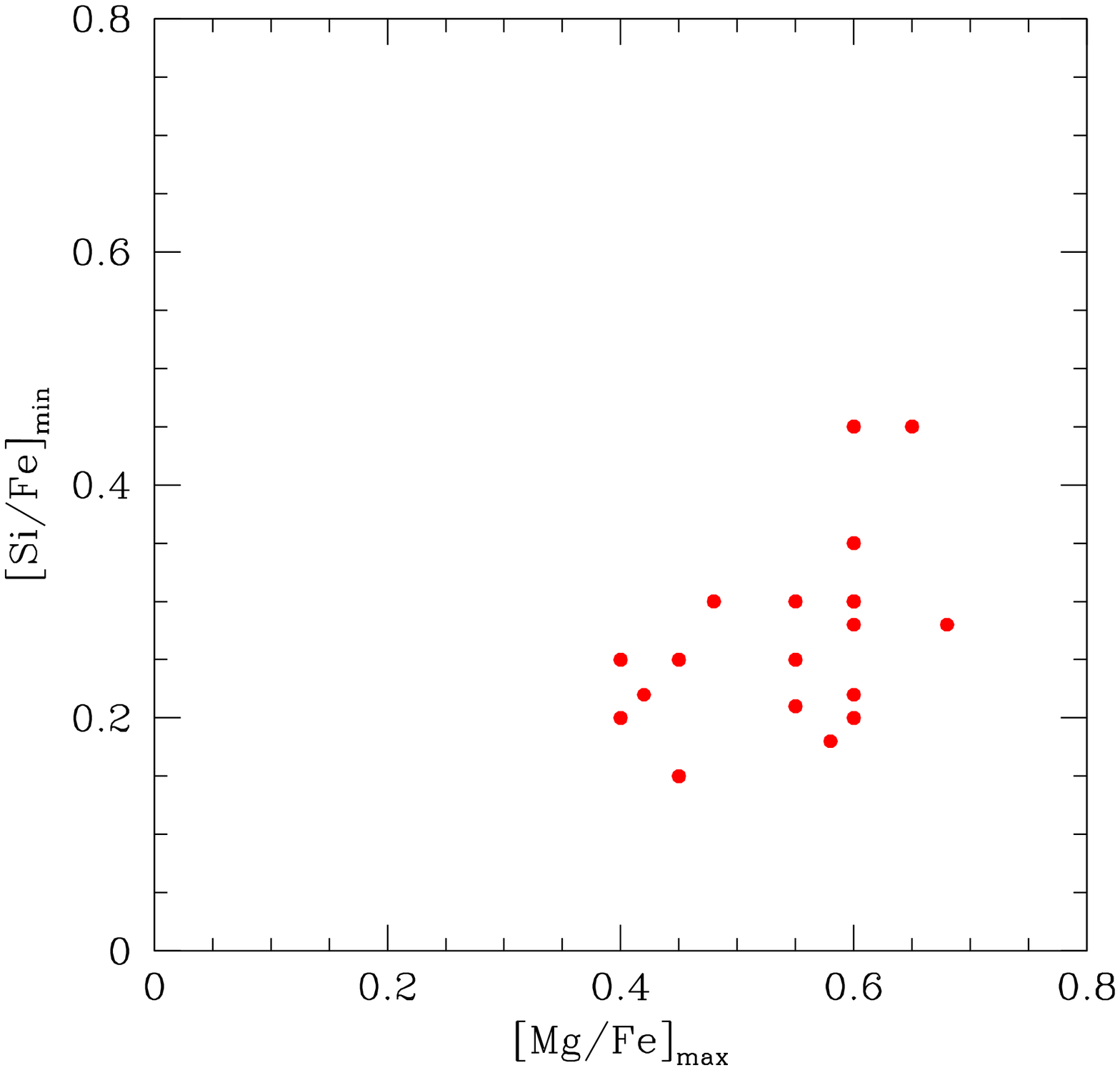}  
\caption{[Mg/Fe]$_{\rm max}$ versus [Si/Fe]$_{\rm min}$ for the 19 GCs.}
\label{f:mgmaxsimin}
\end{figure}

However, these global trends are only marginally instructive. In fact, there are
large differences between individual clusters. Separate plots for each cluster
provide much more insight. Such plots are shown in Figs.~\ref{f:tutteantiu2} and
~\ref{f:tutteantiu3} for the Mg-Al and the Mg-Si relations, respectively.
(In  Fig.~\ref{f:tutteantiu2} NGC~6397 is excluded, since we did not measure  Al
in its stars.) An inspection of these figures reveals a few clear facts:

\begin{itemize} 
\item The maximum Mg and minimum Al and Si abundances, which should be
indicative of the composition of the primordial population (see next
section) may be significantly different in different clusters. Although we
observed only a limited number of stars in each cluster, and in many cases we
were only able to derive upper limits to the Al abundances, there is little
doubt that the minimum Al abundance may vary by almost an order of magnitude
from  cluster to cluster. In the next section we  examine this issue in
more detail, and  try to obtain a better estimate of the minimum Al
abundance for each cluster. On the other hand, maximum Mg and minimum Si
abundances also change from  cluster to cluster. Some cases are obvious, like the
pair of clusters M~5 (NGC~5904) and M~4 (NGC~6121), for which such a 
difference has already been found in
previous analyses (Carney 1996; Ivans et al. 1999, 2001). There is a good
correlation between maximum Mg and minimum  Si abundances derived from
Figure~\ref{f:tutteantiu3}. These values are displayed in
Fig.~\ref{f:mgmaxsimin},  and both are well correlated with the maximum O
abundance observed in each cluster. These fine correlations can be well
explained by different primordial overabundances of the $\alpha-$elements in
the  various clusters. In a separate paper we will study the correlation
existing between the overabundance of $\alpha-$elements and other cluster
parameters. Here, we simply note that it  justifies the global correlation
of Mg and Si abundances. 
\item In a few clusters, none of these elements show significant variations.
Such clusters are generally small and not very metal-poor. Typical cases are
NGC~6121 (M~4), NGC~6171 (M~107), and NGC~288. We notice that all stars in
these  clusters have rather large Al abundances: this finding will be expanded upon in a
future paper. 
\item In the remaining clusters, Al abundances show broad spreads. Clearly,
these variations cannot be explained by observational errors (see 
Table~\ref{t:errabuU}). Interestingly enough, these are also approximately the
same ranges of abundance variations as seen in unevolved cluster stars (Gratton et
al. 2001), pointing to the same origin for the mechanism establishing the
observed chemical pattern. Stars very rich in Al (with [Al/Fe]$\geq 0.8$) show
Mg depletion. This is not surprising,  because Al is about an order of magnitude
less abundant than Mg in the Sun and even more in the primordial stars in GCs. Hence,
when a significant fraction of the  original Mg is transformed into Al (so that
the Mg depletion is detectable),  Al production is comparatively
huge. These Al-very rich and
Mg-depleted stars are present in clusters that are massive (like NGC~2808,
NGC~6388, NGC~6441), quite metal-poor (like NGC~6752), or both (like
NGC~7078=M~15: note that, in this last case, we only have an upper limit to the
Al abundance for the Mg-poor star). Hence, at  variance with the Na-O
anticorrelation, which is present in all GCs, the Mg-Al anticorrelation is
present or particularly prominent only in massive and/or metal-poor GCs. 
\item The most extreme (low) Mg abundances have been detected in three  
NGC~2808 stars. Even these are actual detections, not upper limits. We remind the 
reader that this is one of the only two clusters showing a conspicuous component
of second-generation stars with extreme composition (see Paper VII), together
with NGC~6205 (M~13; Sneden et al. 2004; Cohen and Melendez 2005). Furthermore, it is the
$only$ cluster  where multiple main sequences of quite different He content
(Piotto et al. 2007) have been identified so  far (beside the peculiar case of
$\omega$ Cen, with its multi-peaked metallicity  distribution, see Piotto et al.
2005). 
\end{itemize} 

\begin{figure}
\centering
\includegraphics[scale=0.45]{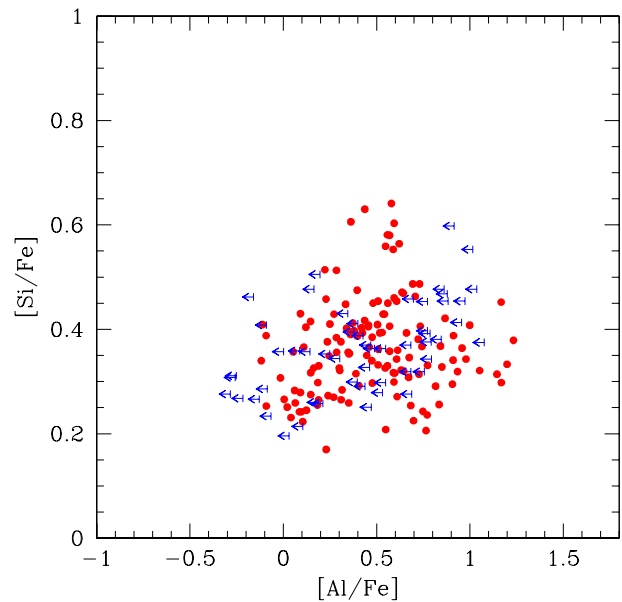}   
\caption{Abundance ratio [Si/Fe] as a function of [Al/Fe] for stars with UVES
spectra in all clusters of this project.}
\label{f:sial}
\end{figure}

We further found that stars with extreme Al overabundance also show Si enhancement
with respect to the remaining stars in the same cluster. Again, this effect is
limited to massive or metal-poor GCs. We are aware of only one previous
detection of the Al-Si correlation in a GC (NGC~6752: Yong et al. 2005). We
now find that the effect is common among GCs, as illustrated by
Fig.~\ref{f:sial}, where we show the correlation resulting by combining 
all stars. We found that the global slope ($0.07\pm0.02$ from 192 stars) is
highly statistically significant. 
Quantification of the amount of Si variation as a function of Al variation in
each individual cluster is hampered by upper limits in Al abundances (see
Fig.\ref{f:tutteantiu2}). However,as Mg and Al are anticorrelated with each
other, we can study the expected anticorrelation between Si and Mg (elements for
which we only have actual detections, no upper limits) to infer information on the $correlation$ between
Si and Al in individual GCs.

\begin{figure}
\centering
\includegraphics[bb=70 153 440 714, clip, scale=0.63]{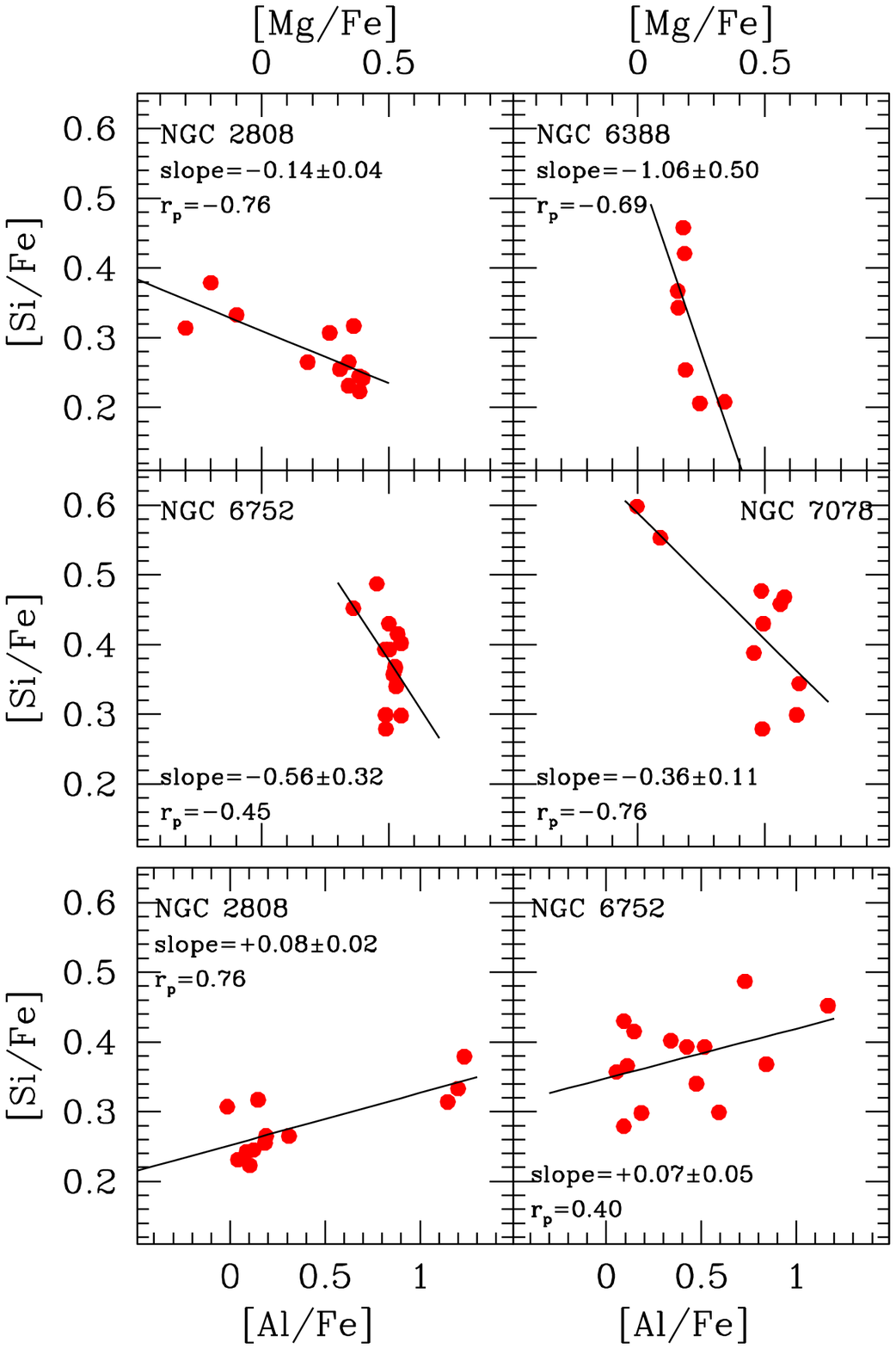}  
\caption{Abundance ratios [Si/Fe] as a function of [Mg/Fe] in four GCs of our
sample with significant slope of the Si-Mg anticorrelation (upper four panels).
Abundance ratio [Si/Fe] as a function of [Al/Fe] for stars in NGC~2808 and
NGC~6752 (bottom two panels).In
each panel we report the slope and the linear regression coefficient of the
correlation.}
\label{f:sial6}
\end{figure}

From data displayed in Fig.~\ref{f:tutteantiu3} we find that slopes
statistically significant (at a level of at least 2$\sigma$) are found 
for NGC~2808, NGC~6388, NGC~6752, and NGC~7078. An enlargement for the Si-Mg
anticorrelation in these clusters, together with the slopes and associated
errors, is shown in the upper part of Fig.~\ref{f:sial6}. In the bottom two
panels of this figure we display the direct Si-Al correlations for the two
clusters with significant slope and no upper limits on Al abundances: NGC~2808
and NGC~6752.

The explanation for  this behaviour is given by Yong et al. (2005) in their
study of giants in NGC~6752. The authors claim that such a correlation is
expected if a ``leakage" from the Mg-Al cycle on $^{28}$Si does occur, and 
discuss how small amounts of $^{29}$Si and $^{30}$Si can even be synthesized
by $neutron-$capture processes in the He-shell of AGB stars. However, the total
abundance  of Si is unlikely to change significantly unless the reaction 
$^{27}$Al($p,\gamma$)$^{28}$Si takes over the reaction
$^{27}$Al($p,\alpha$)$^{24}$Mg  in the Mg-Al chain. The result is that a certain
amount of $^{28}$Si might be produced  in hot bottom burning (HBB) in
intermediate-mass AGB stars by $proton-$capture reactions (see Karakas and 
Lattanzio 2003), giving just the sort of trend displayed in
Fig.~\ref{f:sial6}. 
Incidentally, the variation of Si abundances we found in
NGC~6752 is  exactly the same as that found by Yong et al. (2005).

A strong conclusion that can be drawn from the above results is that only a 
fraction of the polluters responsible for the Na-O anticorrelation actually
produce large quantities of Al. This is likely to be 
related to the large difference
between  the temperature required for the two cycles. The temperature where the
Mg-Al cycle occurs is high enough for a significant production of Si. 
As a result,
the finding of the Si-Al correlation allows us to put a strong constraint on the
temperature at which H-burning occurred in these massive clusters. By  looking
at the temperature dependence of Maxwellian-averaged reaction rates from  NACRE
for proton-captures (right panel of Fig. 8 in Arnould et al. 1999) we can 
evaluate that the reaction producing $^{28}$Si becomes predominant in the Mg-Al 
cycle when the temperature exceeds $T_6 \sim 65$ K. Once again we then confirm 
that the chemical pattern we observe in present-day cluster red giants must  already
be imprinted in the gas by a previous generation of more massive stars. In
currently evolving low-mass stars the CNO processing shell only reaches its 
maximum temperature of about 70 MK very near the tip of the RGB (e.g.
Langer  et al. 1996), while we observe stars still far from this magnitude range
and yet  able to show the Si-Al correlation in their photospheric abundances.

\begin{table}
\centering
\caption[]{Minimum and maximum abundances of Na and Al for the programme clusters from our 
dilution model}
\scriptsize
\begin{tabular}{rcccccc}
\hline
 NGC &  n  &[Na/Fe]$_{\rm min}$ &[Na/Fe]$_{\rm max}$& [Al/Fe]$_{\rm min}$ &[Al/Fe]$_{\rm max}$& rms\\
\hline
 104 & 11 &   0.15 & $0.74\pm 0.06$ &$  0.19\pm 0.11$ & $0.72\pm 0.05$ & 0.08 \\
 288 & 10 &  -0.10 & $0.71\pm 0.18$ &$  0.40\pm 0.02$ & $0.50\pm 0.05$ & 0.06 \\ 
1904 &  8 &  -0.15 & $0.72\pm 0.07$ &$ -0.69\pm 0.24$ & $0.91\pm 0.04$ & 0.16 \\ 
2808 & 12 &  -0.12 & $0.56\pm 0.04$ &$ -0.56\pm 0.17$ & $1.03\pm 0.05$ & 0.17 \\ 
3201 & 13 &  -0.30 & $0.60\pm 0.09$ &$ -0.78\pm 0.24$ & $0.60\pm 0.05$ & 0.18 \\ 
4590 & 12 &  -0.35 & $0.53\pm 0.13$ &$ <0.15	    $ & $<0.65	     $ &      \\ 
5904 & 14 &  -0.25 & $0.60\pm 0.10$ &$ -0.19\pm 0.08$ & $0.61\pm 0.05$ & 0.16 \\ 
6121 & 14 &  -0.05 & $0.74\pm 0.08$ &$  0.59\pm 0.04$ & $0.60\pm 0.09$ & 0.05 \\ 
6171 &  5 &  -0.05 & $0.69\pm 0.07$ &$  0.39\pm 0.10$ & $0.40\pm 0.07$ & 0.07 \\ 
6218 & 11 &  -0.20 & $0.67\pm 0.07$ &$  0.07\pm 0.08$ & $0.35\pm 0.08$ & 0.12 \\ 
6254 & 10 &  -0.30 & $0.56\pm 0.10$ &$ -0.84\pm 0.25$ & $0.76\pm 0.04$ & 0.28 \\ 
6388 &  7 &   0.00 & $0.67\pm 0.05$ &$ -0.33\pm 0.80$ & $0.78\pm 0.08$ & 0.16 \\ 
6397 &    &  -0.35 & $0.71\pm 0.23$ &   	      & 	       &      \\ 
6441 &  5 &  -0.05 & $0.80\pm 0.10$ &$ -0.11\pm 0.23$ & $0.51\pm 0.10$ & 0.06 \\ 
6752 & 14 &  -0.15 & $0.65\pm 0.07$ &$ -0.10\pm 0.09$ & $0.73\pm 0.06$ & 0.16 \\ 
6809 & 14 &  -0.35 & $0.69\pm 0.09$ &$ -0.77\pm 0.39$ & $0.83\pm 0.05$ & 0.19 \\ 
6838 & 12 &   0.00 & $0.76\pm 0.16$ &$  0.24\pm 0.13$ & $0.73\pm 0.06$ & 0.10 \\ 
7078 & 13 &$<-0.05$& $0.70\pm 0.09$ &$< 0.20	    $ &$<0.92	     $ &      \\ 
7099 & 10 &  -0.20 & $0.76\pm 0.14$ &$ -0.55\pm 0.50$ & $1.05\pm 0.14$ & 0.07 \\ 
\hline								        
\end{tabular}
\label{t:dilution}
\end{table}

\section{The Na-Al correlation: tuning a dilution model} 

New insights into the properties of the polluters can be obtained by relating 
the Mg-Al cycle with the Ne-Na cycle. A very fruitful approach is to compare the
production of Na and Al. 

\begin{figure*}
\centering
\includegraphics[bb=10 160 570 600, clip, scale=0.89]{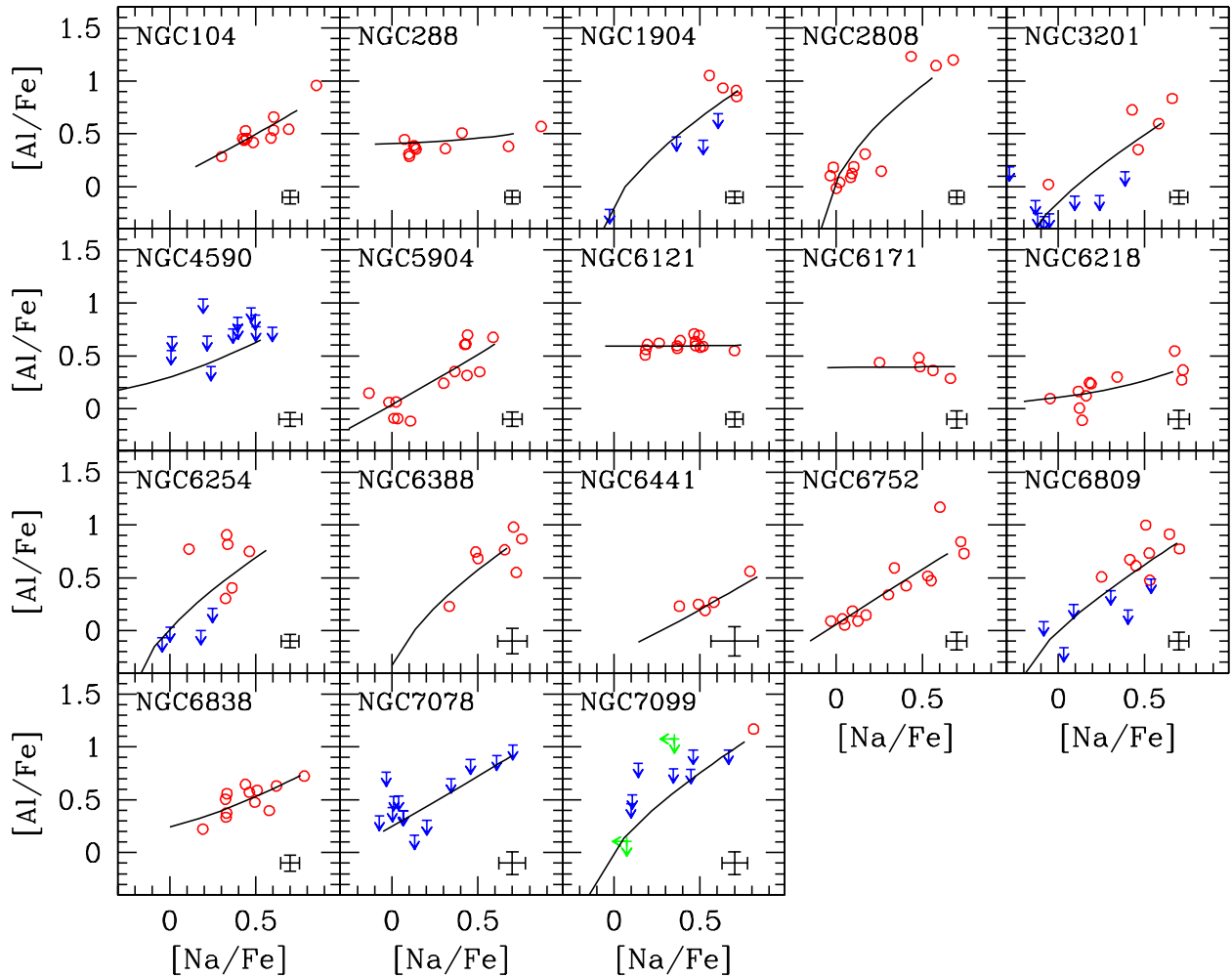}  
\caption{[Al/Fe] ratios as a function of [Na/Fe] ratios for the 18 individual 
GCs, with a dilution model superimposed.}
\label{f:alnaU}
\end{figure*}

We plotted in Fig.~\ref{f:alnaU} the individual runs of Al and Na in each of the
programme clusters. In most cases there is a clear correlation between Al and Na
abundances. However, in spite of the numerous upper limits  for metal-poor or
Al-poor stars, it is also clear that these runs are different from cluster to
cluster. In a few cases there is no appreciable change in the Al abundance, while
Na abundances change by about 0.6 dex (see e.g. the case of NGC~6121=M~4), while in other
clusters the Al abundances change much more than the Na ones. Thus, while the 
Ne-Na and Mg-Al cycles are related, they do not occur in exactly the same
polluting stars, and different clusters have  different proportions of these
different polluters.

To progress in our understanding of the relations between O depletion, 
the Ne-Na, and Mg-Al cycles, we make use of the dilution model considered in
paper VII to explain the Na-O anticorrelation and how it changes from one
cluster to another. A basic assumption of this model is that the polluting
material  has a well-defined composition, and the pattern of abundances in
individual stars  is simply obtained by diluting this polluting material with
variable quantities of primordial material. As mentioned in Paper VII, this
dilution model explains the shape of the Na-O anticorrelation (see also
Prantzos et al. 2007). In this model  the logarithmic abundance of an element
[X] for a given dilution factor $dil$ is given by
\begin{equation}
[X] = \log{[(1-dil)~10^{\rm [Xo]} + dil~10^{\rm [Xp]}]},
\end{equation}
where [Xo] and [Xp] are the logarithmic abundance of the element in the original
and  processed material, respectively. We apply this model to Na and Al
abundances. We notice that a basic property of this model is that there should
be a unique relation between Na and Al abundances in a given cluster. While this
indeed seems the case for most of the clusters, there might be exceptions, the
most notable being NGC~6254, for which there seems to be a genuine scatter of Al
abundances at a given Na abundance. Let us ignore  this difficulty for the
moment. In principle,  [Xo] and [Xp] could be directly derived from observations
for both Na and Al. However,  the sample of stars with UVES spectra is quite
small for a single cluster, and it is very likely that the extremes of the
distributions have not been well sampled. For Na, this concern can be removed by
using the much larger samples of stars with GIRAFFE spectra: for Na we will
adopt the values of [Xo]=[Na/Fe]$_{\rm min}$\ and  [Xp]=[Na/Fe]$_{\rm max}$\
from Table~7 of Paper VII. On the other hand, we can only use the UVES data to
derive [Al/Fe]$_{\rm min}$ and [Al/Fe]$_{\rm max}$. Since we could derive only
upper limits for a number of stars, we applied  a maximum likelihood method (see
Isobe, Feigelson and Nelson 1986) to estimate  [Al/Fe]$_{\rm min}$ and
[Al/Fe]$_{\rm max}$ values, listed in  Table~\ref{t:dilution}.

Dilution sequences obtained by this approach are superimposed on the observed run
of Na and Al abundances in Fig.~\ref{f:alnaU}. They generally do a good job of
fitting the observations. However, there is a large scatter in
the observational data for some clusters (e.g. NGC~6254).
We think that there are a few cases where
our dilution model is not fully compatible with observations, most likely
because we cannot assume a single composition for the polluters.

\begin{figure}
\centering
\includegraphics[scale=0.45]{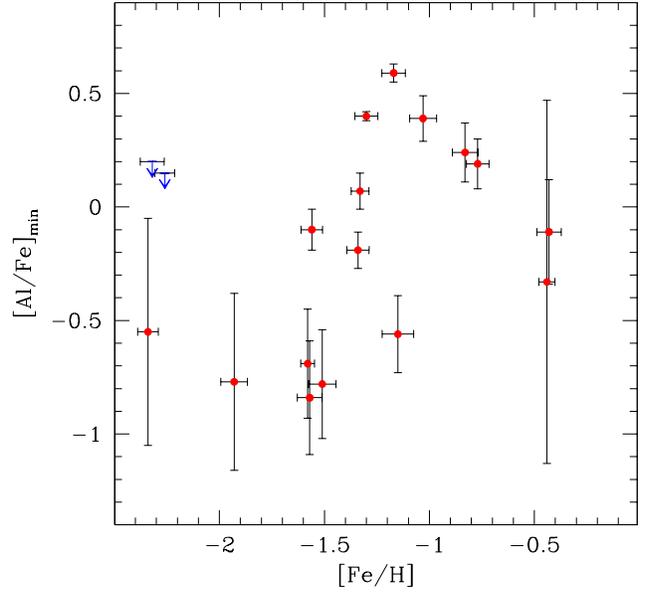}
\caption{[Al/Fe]$_{\rm min}$ versus [Fe/H] for the 19 GCs; the two upper limits
are for NGC~4590 and NGC~7078.}
\label{f:alfe}
\end{figure}

In our model, [Al/Fe]$_{\rm min}$ represents the original Al abundance in the
cluster. We plotted the run of [Al/Fe]$_{\rm min}$ with [Fe/H] in
Fig.~\ref{f:alfe}. We obtain low values of Al abundances in the most metal-poor
clusters: in a couple of cases they are upper limits, because we did not detect Al lines
in these clusters. Metal-rich clusters usually
have very high values of [Al/Fe]$_{\rm min}$. On the other hand, we find a
huge range in Al abundances in clusters of intermediate metallicity. We will
re-examine this point in a forthcoming paper, but only anticipate here that most
of this scatter can be explained  by the origin of these clusters (disc or inner
halo versus outer halo clusters).

\begin{figure}
\centering
\includegraphics[scale=0.45]{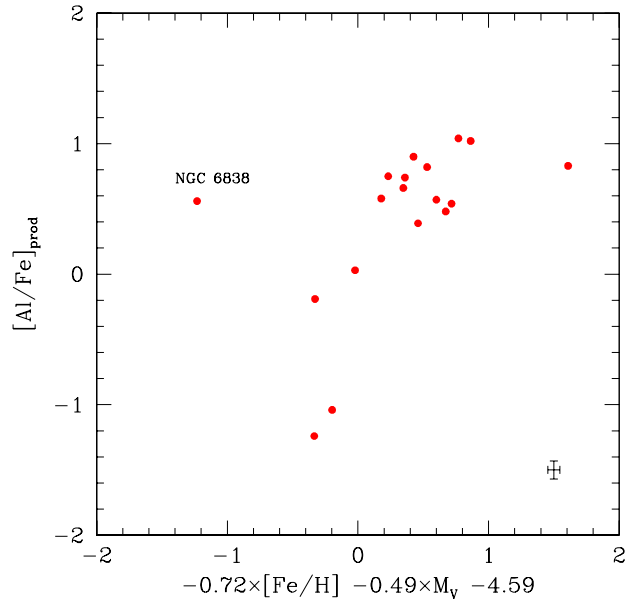}  
\caption{Run of [Al/Fe]$_{\rm prod}$ for the GC of our sample as a function of a
linear combination of metallicity [Fe/H] and cluster luminosity $M_V$.  The
relation was computed excluding NGC~6838.}
\label{f:alprod}
\end{figure}

On the other hand, [Al/Fe]$_{\rm max}$, or better the amount of Al
produced\footnote{ [Al/Fe]$_{\rm prod}=\log{(10^{\rm [Al/Fe]_{ max}}-10^{\rm
[Al/Fe]_{min}})}$},  depends on the yields of typical polluters. It is then
interesting to notice that we found [Al/Fe]$_{\rm prod}$\ to be 
correlated with a linear combination of metallicity and cluster luminosity, as
shown in Fig.~\ref{f:alprod}. This is true for all GCs in our sample except
NGC~6838, which seems to have too large an Al production with respect to its
metallicity and total mass. We have at present no explanation for this cluster
so we excluded it from the fit. We note that this is  the same combination of
cluster metallicity and luminosity  that we obtained from a multivariate analysis
of  [O/Fe]$_{\rm min}$\ in Paper VII (see Fig. 20 there).  The implication is
that the polluters that produced Al are actually the same as destroyed O and
produced Na, but that their properties change regularly with cluster luminosity
and metallicity.

The dependence on metallicity is not surprising, since HBB is expected to occur
at a higher temperature in more metal-poor stars (see Lattanzio, Forestini and
Charbonnel 2000).
These stars can produce Al and destroy O more efficiently. On the  other hand,
cluster luminosity is probably a proxy for the cluster mass,  and the latter might
be a proxy for polluter mass. This may be understood within the context that
dynamical evolution of clusters occurs more rapidly in clusters of high mass
(and luminosity). In this case, we should expect that the typical polluter mass
is higher in more massive (and luminous) clusters. The trend towards increasing
[Al/Fe]$_{\rm max}$ and decreasing [Na/Fe]$_{\rm max}$ and [O/Fe]$_{\rm min}$
with increasing cluster luminosity should be a consequence of a similar trend
with increasing typical polluter mass. 

We may compare this result with the very recent nucleosynthetic yields for AGB
polluters by Ventura \& D'Antona. (2009). We indeed find similar trends over the
mass range  4-6~M$_\odot$ (the mass range changes somewhat with metallicity, the
value we give  is for $Z=10^{-3}$); however, the agreement is only qualitative.
In the models Al production and O destruction are not enough, and/or Na
production  is much too sensitive to mass. As a consequence, even playing with
the minimum and maximum  mass of the polluters, we are never able to reproduce
the abundances of all of these  elements in any cluster. For instance, in the case
of NGC~2808, we may obtain a reasonably good production of Na and Al; adopting a
mass range for the polluters between 4.5 and 6.3~M$_\odot$ (and a Salpeter
initial mass function -but this is not a critical  assumption, given the 
narrow
range of masses involved) we obtain [Na/Fe]$\sim 0.6$ and [Al/Fe]$\sim 1.0$, to
be compared with [Na/Fe]$_{\rm max}$=0.56 and [Al/Fe]$_{\rm max}$=0.92. However,
we get [O/Fe]$\sim -0.1$, which is much larger than the observed value of
[O/Fe]$_{\rm min}=-1$ (see Paper VII).  In the case of NGC~6121, a slightly 
wider
mass range of the polluters (between 4.2 and 6.3~M$_\odot$) again reproduces
the abundances of O and Na quite well ([O/Fe]$\sim 0$ and [Na/Fe]$\sim 0.75$, to
be compared with [O/Fe]$_{\rm min}=-0.2$ and  [Na/Fe]$_{\rm max}$=0.74), but it
overestimates the production of Al ([Al/Fe]$\sim 1$ against [Al/Fe]$_{\rm
max}$=0.65). We conclude that more improvement in the models is required for
such a comparison to be meaningful.

We remind the reader that there are a few clusters for which a unique dilution sequence
does not reproduce the run of [Al/Fe] with [Na/Fe]. Examples are NGC~3201,
NGC~6254, and NGC~6809, but there are other more dubious cases, like  NGC~5904.
This result is not new, since Johnson et al. (2005) find a similar  result for
NGC~5272 (M~3). Such a result would suggest that in these clusters there were more
generations of stars, or perhaps even a continuum of second generations.  This
last hypothesis would agree better with the massive star polluters, where  each
polluter might have its own progeny.

Let us then consider the models for fast-rotating massive stars. We may use the
yields given in Table 5 of Decressin et al. (2007), who give the predicted
composition of the wind as a function of stellar mass. These models predict a
roughly constant Na production and a production of Al and destruction of O that
are an increasing function of the mass. This might indeed agree with
observations. However, a closer look reveals several inconsistencies. First, Al
production and O destruction change much less with mass than do the observed
cluster to cluster variations in the maximum Al and minimum O abundances. This
discrepancy becomes even worse if we consider that stars over  quite  wide mass
ranges should be taken into account to provide enough mass for the second
generation. A range in stellar rotation rates is no help, as it yields a range
in Na abundances, which is not observed. This is the same difficulty as was
found for the AGB model, and it may point toward basic problems in the relevant
cross sections.

The rotating massive star scenario has an additional difficulty. In
fact, in this scheme, material expelled from the polluting stars is not mixed
into a single star-forming cloud, but rather second generation stars form
around each of the fast-rotating, polluting massive stars. We then expect
different O, Na, and Al yields, depending on the mass and rotation rate of the
polluters. If we combine these different yields with a range in dilution
factors, we expect that second-generation stars fill a triangle in the [Na/Fe]
versus [Al/Fe] plane, the vertices being defined by the original [Na/Fe] and [Al/Fe]
values, and by the [Na/Fe] and [Al/Fe] values given by the winds of the most and
less massive fast-rotating star, respectively. The closer these last two values,
the less is the mass range allowed for polluters. While a range in polluter
masses seem indeed present in clusters like NGC~3201 or NGC~6254 
(or NGC~5272=M~3: see
Sneden et al. 2004), the sequences seem almost monoparametric in other clusters,
including NGC~104 (47 Tuc), NGC~2808, NGC~6752 (see also Yong et al. 2003), or 
NGC~6205 (M~13, see Sneden et al. 2004).  More extensive samples of stars 
with  accurate
determinations of the Al abundances are required to clarify this issue.

\section{Summary}

In this paper, we have derived atmospheric parameters, abundances of 
Fe and of the elements involved in proton capture at high temperatures (O,
Na, Mg, Al, and Si) in 202 stars in 17 GCs. This analysis
is based on FLAMES-UVES spectra. Adding the two clusters with similar
data whose analysis has already been presented in previous papers (NGC~6441:
Paper III; and NGC~6388: Paper VI), we now have a large data set (214 stars)
from 19 GCs. This material complements results derived from the
FLAMES-GIRAFFE spectra of about 2,000 stars, as discussed in Paper VII,
providing abundance analysis for a sample of GC stars of unprecedented size.

The analysis of this huge dataset was done as automatically as possible.
We put great care to reducing internal errors. Effective temperature were
generally derived exploiting the $T_{\rm eff}-K$\ magnitude relation
within each cluster, calibrated using the relation between
$T_{\rm eff}$ and $(V-K)$\ colours. This allows us to minimise the impact of
differential reddening (when present), and to reduce star-to-star errors
in the effective temperatures to a few K. This was crucial for enabling us 
to discuss star-to-star abundance differences within each cluster.

Data mining of this huge dataset is in progress. In the present paper, we
focused on the star-to-star scatter in the abundances of O, Na, Mg, Al,
and Si. This was very important for discussing the relation between the Ne-Na
and Mg-Al cycles. The main results we obtained are as follows:
\begin{itemize}
\item We confirm the very small scatter in Fe abundances found in Paper
VII. Regardless of the mechanism responsible for the variations in the
abundances of O, Na, Mg, Al, and Si, it does not produce
significant amounts of Fe.
\item As in paper VII, we found variations in O and Na abundances
(anticorrelated with each other) in all surveyed clusters, including small
ones like NGC~288, NGC~6397, or NGC~6838 (M~71). While minimum O abundances
change greatly from cluster to cluster, the maximum Na abundances show
only a narrow spread. Clusters like NGC~288, NGC~6121 (M~4), and NGC~6171
(M~107) show no variation in the abundances of Al at all. This contrasts
with the huge spread in the Al abundances found in other
clusters, like NGC~2808, NGC~3201, NGC~6752. This clearly indicates that only
a subset of the polluters responsible for the variations of the O and Na
abundances are responsible for those of the Al abundances. That is, 
only a
fraction of the polluting stars reached temperatures high enough for the Mg-Al
cycle to be active.
\item Significant variations in Mg and Si are seen whenever large
overabundances of Al are observed. While the Mg depletion was expected in
these cases, the observed overabundances of Si indicate that the ``leakage"
mechanism from the Mg-Al cycle proposed by Yong et al. (2005) to explain
their data for NGC~6752, is in fact active also in other clusters. This
requires temperatures above 65 MK for the Mg-Al cycle.
\item We used a dilution model to derive minimum and maximum Al abundances
from the Na-Al relation drawn from our UVES data, and minimum and maximum
Na abundances obtained from the much more extensive GIRAFFE sample. These
should reflect the primordial and polluter Al abundances, respectively. The
peculiar run of the minimum Al abundances will be discussed in a
forthcoming paper. However, from maximum and minimum Al abundances, we
were able to derive the Al produced within the polluters. This was found to
be correlated with the same combination of cluster luminosity (likely
a proxy for the mass) and metallicity that we obtained from an examination 
of the
minimum O abundances in Paper VII. This indicates that Al production and O
destruction are closely related, possibly indicative of a
common dependence on the typical mass of the polluters.
\item A comparison of these results with current nucleosynthesis
predictions from either massive AGB stars or fast-rotating massive stars
shows that in both cases a qualitative, but not quantitative agreement can
be obtained. Further theoretical work is clearly required.
\item Finally, we found that our dilution model satisfactorily explains 
most of the clusters. However, the spread in Al abundances at a given Na
abundance found in a few clusters (the clearer example being NGC~6254=M~10)
shows that in some cases there was more than a single pool of gas from
which second generation stars formed. We argue that a careful and
extensive study of the Na-Al relation may provide crucial information on
the nature of the polluters.
\end{itemize}

\begin{acknowledgements}
We wish to thank the ESO Service Mode personnel for their work, F. D'Antona for
discussions, V. D'Orazi for her help, and the referee, Mike Bessell, for helping
us to improve the paper.
This publication makes use of data products from the Two Micron All Sky Survey,
which is a joint project of the University of Massachusetts and the Infrared
Processing and Analysis Center/California Institute of Technology, funded by the
National Aeronautics and Space Administration and the National Science
Foundation. This research has made use of the SIMBAD
database, operated at CDS, Strasbourg, France and of NASA's Astrophysical Data
System.
This work was partially funded by the Italian MIUR under PRIN
2003029437. We also acknowledge partial support from the grant INAF 2005
``Experimenting nucleosynthesis in clean environments".  SL is grateful to the
DFG cluster of excellence ``Origin and Structure of the Universe" for support. 
\end{acknowledgements}


\begin{thebibliography}{}

\bibitem[]{}
 Alonso, A., Arribas, S., Martinez-Roger, C. 1999, A\&AS, 140, 261 

\bibitem[]{}
Arnould, M., Goriely, S., Jorissen, A. 1999, A\&A, 347, 572  

\bibitem[]{} Bekki, K. 2006, MNRAS, 367, L24 

\bibitem[]{}
 Bragaglia, A., Carretta, E., Gratton, R.G. et al. 2001, AJ, 121, 327
 
\bibitem[]{} Carney, B.W. 1996, PASP, 108, 900

\bibitem[]{}
 Carretta, E., Bragaglia, A., Gratton R.G., Leone, F., Recio-Blanco,
 A., Lucatello, S. 2006, A\&A, 450, 523 (Paper I)

\bibitem[]{}
 Carretta, E., Bragaglia, A., Gratton R.G., Lucatello, S., \&
 Momany, Y. 2007a, A\&A, 464, 927  (Paper II) 

\bibitem[]{} Carretta, E., Cohen, 
J.~G., Gratton, R.~G., Behr, B.~B. 2001, AJ, 122, 1469 

\bibitem[]{} Carretta, E. et al. 2007b, A\&A, 464, 967 (Paper VI) 

\bibitem[]{} Carretta, E. et al. 2007c,  A\&A, 464, 939 (Paper IV) 

\bibitem[]{} Carretta, E. et al. 2009,  A\&A, in press (Paper VII) 

\bibitem[]{} Cayrel, R. 1986, A\&A, 168, 8

\bibitem[]{} Charbonnel, C., Prantzos, N. 2006, in Globular Clusters, Guide to
Galaxies, ed. T. Richtler, et al, Un. Concepcion (arXiv:astro-ph/0606220v1)

\bibitem[]{} Cohen, J.G., Melendez, J. 2005, AJ, 129, 303 

\bibitem[]{} Cohen, J.G., Briley, M.M., Stetson, P.B. 2002, AJ, 123, 2525 

\bibitem[]{} Cohen, J.~G., Gratton, R.~G., Behr, B.~B.,  Carretta, E. 1999, 
ApJ, 523, 739 

\bibitem[]{} Decressin, T., Meynet, G., Charbonnel C. Prantzos, N.,
Ekstrom, S. 2007, A\&A, 464, 1029 

\bibitem[]{} Denisenkov, P.A., Denisenkova, S.N. 1989, A.Tsir., 1538, 11

\bibitem[]{} Fulbright, J.P. 2000, AJ, 120, 1841

\bibitem[]{} Fulbright,J.P., McWilliam, A., Rich M.R. 2007, ApJ, 661, 1152

\bibitem[]{} Gehren, T., Shi, J.R., Zhang, H.W., Zhao, G., Korn, A.J. 2006,
 A\&A, 451, 1065

\bibitem[]{} Gratton, R.G., Bonifacio, P., Bragaglia, A., et al. 2001, A\&A, 369, 87

\bibitem[]{}
 Gratton, R.G., Carretta, E., Claudi, R., Lucatello, S., 
 Barbieri, M. 2003, A\&A, 404, 187

\bibitem[]{} Gratton, R.G., Carretta, E., Eriksson, K.,  Gustafsson, B. 1999,
 A\&A, 350, 955 

\bibitem[]{} Gratton, R.G., Lucatello, S., Bragaglia, A., Carretta, E., Momany,
Y., Pancino, E., Valenti, E. 2006,  A\&A, 455, 271 (Paper III) 

\bibitem[]{} Gratton, R.G., Sneden, C., \& Carretta, E. 2004, ARA\&A, 42, 385

\bibitem[]{} 
Gratton, R.G., Sneden, C., Carretta, E., \& Bragaglia, A. 2000, A\&A, 354, 169

\bibitem[]{} Gratton, R.G. et al. 2007, A\&A, 464, 953 (Paper V) 

\bibitem[]{} Harris, W.~E. 1996, AJ, 112, 1487 

\bibitem[]{} Isobe, T., Feigelson, E.D., Nelson, P.I. 1986, ApJ, 306, 490

\bibitem[]{} Ivans, I.~I., Kraft,  R.~P., Sneden, C., Smith, G.~H., Rich, R.~M.,
Shetrone, M. 2001, AJ, 122, 1438 

\bibitem[]{} Ivans, I.~I., Sneden, C.,  Kraft, R.~P., Suntzeff, N.~B., Smith,
V.~V., Langer, G.~E.,  Fulbright, J.~P. 1999, AJ, 118, 1273 

\bibitem[]{} Johnson, C.I., Kraft, R.P., Pilachowski, C.A., Sneden, C., Ivans, 
I.I., Benman, G. 2005, PASP, 117, 1308

\bibitem[]{} Johnson, J. 2002, ApJS, 139, 219

\bibitem[]{} Jonsell, K., Edvardsson, B., Gustaffsson, B., Magain, P., Nissen,
  P.E., Asplund, M. 2005, A\&A, 440, 321

\bibitem[]{} Karakas, A., Lattanzio, J.C. 2003, PASA, 20, 279

\bibitem[]{} Kraft, R.P. 1994, PASP, 106, 553

\bibitem[]{} Kurucz, R.L. 1993, CD-ROM 13, Smithsonian Astrophysical
 
\bibitem[]{} Langer, G.E., Hoffman, R., \& Sneden, C. 1993, PASP, 105, 301

\bibitem[]{} Langer, G.E., Hoffman, R., Zaidins, C.S.. 1996, PASP, 109, 244

\bibitem[]{} Lattanzio, J.C., Forestini, M., Charbonnel, C. 2000, MemSAIt, 71,
 737 

\bibitem[]{} Mackey, A.D., van den Bergh, S. 2005, MNRAS, 360, 631

\bibitem[]{} Magain, P. 1984, A\&A, 134, 189

\bibitem[]{} Parmentier, G., Gilmore, G. 2001, A\&A, 378, 97

\bibitem[]{} Pasquini, L., et al. 2002, The Messenger, 110, 1

\bibitem[]{} Piotto et al. 2005, ApJ, 621, 777 

\bibitem[]{} Piotto et al. 2007, ApJ, 661, L53 

\bibitem[]{} Prantzos, N., Charbonnel, C., 2006, A\&A, 458, 135

\bibitem[]{} Prantzos, N., Charbonnel, C., Iliadis, C. 2007, A\&A, 470, 179 

\bibitem[]{} Reddy, B.E., Tomkin, J., Lambert, D.L., Allende prieto, C. 2003,
  MNRAS, 340, 304

\bibitem[]{} Skrutskie, M.F., et al. 2006, AJ, 131, 1163
 
\bibitem[]{} 
 Sneden, C., Kraft, R.P., Guhathakurta, P., Peterson, R.C., Fulbright, J.P. 
 2004, AJ, 127, 2162 

\bibitem[]{} Venn, K.A., Irwin, M., Shetrone, M.D., Tout, C.A., Hill, V.,
Tolstoy, E. 2004, AJ, 128, 1177

\bibitem[]{} 
 Ventura, P., D'Antona, F. 2009, A\&A, 499, 835

\bibitem[]{} 
 Ventura, P., D'Antona, F., Mazzitelli, I., Gratton, R. 2001, ApJ, 550, L65

\bibitem[]{} Yong, D., Grundahl, F., Lambert, D.L., Nissen, P.E., Shetrone, M.D.
 2003, A\&A,402, 985 

\bibitem[]{} Yong, D., Grundahl, F.,  Nissen, P.E., Jensen, H.R., 
 Lambert, D.L. 2005, A\&A, 438, 875

\end{thebibliography}
\end{document}